\documentclass[acmsmall]{acmart}

\usepackage{multirow}

\AtBeginDocument{%
  \providecommand\BibTeX{{%
    \normalfont B\kern-0.5em{\scshape i\kern-0.25em b}\kern-0.8em\TeX}}}

\setcopyright{acmcopyright}
\copyrightyear{2020}
\acmYear{2020}

\acmJournal{TOCS}

\begin{document}

\title{MANA: Microarchitecting an Instruction Prefetcher}

\author{Ali Ansari}
\email{ali.ansari@epfl.ch}
\affiliation{%
  \institution{EPFL}
  \city{Lausanne}
  \country{Switzerland}
  \postcode{1015}
}

\author{Fatemeh Golshan}
\affiliation{%
  \institution{Sharif University of Technology}
  \streetaddress{Azadi Avenue}
  \city{Tehran}
  \country{Iran}}
\email{fgolshan@ce.sharif.edu}

\author{Pejman Lotfi-Kamran}
\affiliation{%
	\institution{Institute for Research in Fundamental Sciences (IPM)}
  \city{Tehran}
  \country{Iran}
}
\email{plotfi@ipm.ir}

\author{Hamid Sarbazi-Azad}
\affiliation{%
 \institution{Sharif University of Technology}
 \streetaddress{Azadi Avenue}
 \city{Tehran}
 \country{Iran}}
\email{azad@sharif.edu}
\affiliation{%
	\institution{Institute for Research in Fundamental Sciences (IPM)}
  \city{Tehran}
  \country{Iran}
  }
\email{azad@ipm.ir}

\renewcommand{\shortauthors}{Ansari, et al.}

\begin{abstract}
	L1 instruction (L1-I) cache misses are a source of performance bottleneck. Sequential prefetchers are simple solutions to mitigate this problem; however, prior work has shown that these prefetchers leave considerable potentials uncovered. This observation has motivated many researchers to come up with more advanced instruction prefetchers. In 2011, Proactive Instruction Fetch (PIF) showed that a hardware prefetcher could effectively eliminate all of the instruction-cache misses. However, its enormous storage cost makes it an impractical solution. Consequently, reducing the storage cost was the main research focus in the instruction prefetching in the past decade.

Several instruction prefetchers, including RDIP and Shotgun, were proposed to offer PIF-level performance with significantly lower storage overhead. However, our findings show that there is a considerable performance gap between these proposals and PIF. While these proposals use different mechanisms for instruction prefetching, the performance gap is largely not because of the mechanism, and instead, is due to not having sufficient storage. Prior proposals suffer from one or both of the following shortcomings: (1) a large number of metadata records to cover the potential, and (2) a high storage cost of each record. The first problem causes metadata-miss, and the second problem prohibits the prefetcher from storing enough records within reasonably-sized storage.

In this paper, we make the case that the key to designing a powerful and cost-effective instruction prefetcher is choosing the right metadata record and microarchitecting the prefetcher to minimize the storage.   We find that high spatial correlation among instruction accesses leads to compact, accurate, and minimal metadata records. We also show that chaining these records is an effective way to enable robust and timely prefetching. Based on the findings, we propose \textit{MANA}, which offers PIF-level performance with 15.7$\times$ lower storage cost. MANA outperforms RDIP and Shotgun by 12.5 and 29\%, respectively. We also evaluate a version of MANA with no storage overhead and show that it offers 98\% of the peak performance benefits.

\end{abstract}



\maketitle

       \section{Introduction}

Instruction cache misses are a well-known source of performance degradation when the limited-capacity L1 instruction (L1-I) cache cannot capture a large number of instruction blocks\footnote{By instruction block, we mean a 64-byte cache line.} demanded by a processor~\cite{ayers:memory, ferdman:clearing, hardavellas:database, kanev:profiling, lim:understanding, bakhshalipour2018domino, bakhshalipour2019evaluation, bakhshalipour2018fast,}. In modern processors, the address generator is responsible for filling the~\textit{fetch queue}, which is a queue of addresses expected to be demanded by the processor. The fetch engine looks up the L1-I cache to extract the instructions of the addresses in the fetch queue. These instructions are decoded and sent to the core backend for execution. While modern processors support out-of-order execution, instruction supply to the core backend is still in-order. Therefore, if the instruction at the head of the fetch queue misses in the L1-I cache, the core will no longer be fed by new instructions until the missing instruction arrives at the L1-I cache, which results in performance degradation. However, the fetch engine may continue to fetch the remaining addresses in the fetch queue from the L1-I.

Instruction prefetching is a technique to address this problem. An instruction prefetcher predicts the future cache misses and sends prefetch requests to fill the cache before demand requests arrive. The most common instruction prefetchers are sequential prefetchers that, upon activation, send prefetch requests for a few subsequent blocks~\cite{smith:sequential, xia:instruction}. While sequential prefetchers are used in commodity processors~\cite{jouppi1990improving, ramirez:fetching, santana:enlarging, smith:sequential}, prior work has shown that such prefetchers leave a significant fraction of instruction misses uncovered, and hence, there is a substantial opportunity for improvement~\cite{kolli2013rdip, ferdman:pif, kumar:boomerang}. 

Sequential prefetchers' limitations motivated researchers to propose more sophisticated prefetchers. Proactive Instruction Fetch (PIF) is a pioneer that showed a hardware instruction prefetcher could eliminate most of the instruction cache misses~\cite{ferdman:pif}. However, the proposed prefetcher is impractical because of its high storage cost. Nonetheless, PIF motivated many researchers to develop effective and storage-efficient prefetchers~\cite{kaynak:shift, kolli2013rdip, kumar:boomerang, kumar:shotgun, ayers2019asmdb, kallurkar:ptask}.

We evaluate Return-Address-Stack Directed Instruction Prefetcher (RDIP)~\cite{kolli2013rdip}, Shotgun~\cite{kumar:shotgun}, and PIF, as the three state-of-the-art prefetchers that use entirely different approaches for instruction prefetching. We show that PIF offers considerably higher speedup as compared to the other two prefetchers. Our findings indicate that the inefficiency of the competitors is mainly because of the \textit{metadata-miss} as a result of not having enough storage. If the storage budget is unlimited, they offer almost the same level of performance. These results suggest that designing a strong instruction prefetcher is mainly about the storage-efficiency of the prefetcher.


In this paper, we argue that designing a strong instruction prefetcher needs considering the following. (1) Prefetchers create and store metadata records and prefetch accordingly. These metadata records should carefully be chosen to minimize the number of distinct records. The small number of such records enables a prefetcher to experience a high hit ratio in its metadata storage. (2) Along with the small number of distinct records, each record should need as few bits as possible. This feature enables a prefetcher to store a larger number of records when a specific storage budget is provided.

Based on the guideline, we introduce \textit{MANA} prefetcher, which benefits from spatial correlation. Not only spatial correlation offers compact metadata records, but also the number of distinct records is small. We also find that chaining spatial-metadata records provides a space-efficient way to take advantage of temporal correlation among spatial records to maximize the benefit. We organize the metadata storage so that MANA stores as few records as possible, and each record requires a minimal storage cost. The low storage cost enables MANA to achieve over 92\% of the performance potential with only 15~KB of storage, considerably outperforming RDIP and Shotgun. Moreover, MANA can prefetch for smaller L1-I caches to eliminate the storage overhead as compared to the baseline design.

        \section{Background}

This section introduces primary instruction prefetchers.

\subsection{Temporal Prefetchers}

Temporal prefetching is based on the fact that the sequence of instruction cache accesses or misses is repetitive, and hence, predictable~\cite{ferdman:tifs, ferdman:pif}. Consequently, temporal instruction prefetchers record and replay the sequence to eliminate future instruction cache misses. Temporal Instruction Fetch Streaming (TIFS)~\cite{ferdman:tifs} records and replays the sequence of misses and offers adequately good results. However, PIF~\cite{ferdman:pif} offers a more significant improvement as compared to TIFS by recording and replaying the sequence of instruction accesses.

Temporal prefetchers have two main components: a history in which the sequence is recorded, and an index table that determines the last location of every address (more precisely trigger address as we discuss shortly) in the history. Such a structure imposes a high storage cost, which is the main shortcoming of temporal prefetchers~\cite{bakhshalipour2017efficient, bakhshalipour2019bingo}. As an example, PIF requires more than 200~KB storage budget per core to work effectively. As providing such ample storage is not feasible, researchers proposed techniques to reduce the storage cost. 

Shared History Instruction Fetch (SHIFT)~\cite{kaynak:shift} shares PIF's metadata among cores and virtualizes~\cite{burcea:predictor} it in the last-level cache (LLC). In multi-core processors, when cores execute the same application, the sequence that is created by one core can be used for others as well. As a result, it is not necessary to use a dedicated metadata table for each core. Sharing and virtualizing PIF's metadata in the LLC reduced prefetcher's storage cost from more than 200~KB per core to a total of 240~KB virtualized in an 8 MB LLC\footnote{SHIFT's storage cost is proportional to the LLC size.}. However, the results show that sharing and virtualizing the metadata in the LLC degrades the performance boost of SHIFT as compared to PIF.

\subsection{RAS Directed Instruction Prefetcher (RDIP)}

Return-Address-Stack Directed Instruction Prefetcher (RDIP)~\cite{kolli2013rdip} is proposed to offer PIF-level performance with a significantly lower storage cost. RDIP observes that the current state of the return-address-stack (RAS) can give an accurate representation of the program's state. To exploit this observation, RDIP XORs the four top entries of the RAS and calls it a \textit{signature}. Then it assigns the observed instruction cache misses to the corresponding signature. Finally, it stores these misses in a set-associative table that is looked up using the signature. RDIP reduces the per-core storage to over 60~KB. While RDIP requires considerably lower storage as compared to PIF, it still needs a significant storage budget. 

\subsection{BTB-Directed Prefetchers}

Branch Target Buffer (BTB)-directed prefetchers are advertised as metadata-free prefetchers. Fetch Directed Instruction Prefetcher (FDIP)~\cite{reinman:fetch} is the pioneer of such prefetchers. The main idea is to decouple the fetch engine from the branch predictor unit. This way, the branch predictor unit goes ahead of the fetch stream to discover the instruction blocks that will be demanded shortly. The prefetcher checks if any of those blocks are missing and prefetches them. For this goal, BTB-directed prefetchers use a deep queue of discovered instructions named Fetch Target Queue (FTQ). FTQ is used to fill the gap between the fetch engine and the branch prediction unit. The prefetcher makes progress instruction by instruction, finds branch instructions by looking up the BTB, consults the branch predictor to determine their target, and inserts the instructions into the FTQ. The instructions at the head of the FTQ are the demand instructions. The remaining entries, on the other hand, enable the prefetch engine to look up the cache and prefetches the missing ones.   

The main bottleneck of BTB-directed prefetchers is BTB misses~\cite{kumar:boomerang, kumar:shotgun}. To correctly go far ahead of the fetch stream, such a prefetcher needs to detect the branches and predict their target. BTB is the component that is used to detect branch instructions. The branches' directions can be identified using a branch predictor. Kumar et al. investigated the effect of these two components on BTB-directed instruction prefetching~\cite{kumar:boomerang}. They showed that while the branch predictor's accuracy is not important, BTB misses can significantly limit BTB-directed prefetchers' efficiency. Hence, their proposal, Boomerang, not only prefetches for the L1-I caches but also the BTB.

Boomerang uses a basic block-oriented BTB. The main advantage of this BTB type is that the target of each branch is the starting address of a basic block. As a result, Boomerang can detect BTB misses. Moreover, Boomerang uses an instruction pre-decoder to detect branch instructions and extract their targets. By detecting the BTB misses and extracting them from the instruction blocks, Boomerang can prefill BTB to continue going ahead of the fetch stream.

With Boomerang, BTB misses are still a bottleneck~\cite{kumar:shotgun}. Boomerang stalls on a BTB miss and waits until it is resolved. However, resolving a BTB miss requires the instruction block that holds the missing BTB entry to be present in the cache. Otherwise, Boomerang should fetch that block, and then, pre-decode it to extract the required BTB entry and fill in the BTB. As fetching an instruction block takes considerable clock cycles, in workloads with very large instruction footprints in which instruction block and BTB misses are frequent, Boomerang does not offer a considerable performance boost~\cite{kumar:shotgun}. 

To address this problem, Kumar et al. proposed a new BTB organization within Shotgun prefetcher~\cite{kumar:shotgun} to offer two advantages on top of Boomerang. First, Shotgun prefetches for L1-I without the need to hold all the basic blocks of the instruction cache blocks in the BTB. Moreover, the new BTB covers a larger address space than conventional BTB's. Shotgun has three distinct BTB structures, one for unconditional branches (U-BTB), the other one for conditional branches (C-BTB), and the last one for function and trap returns (RIB). The idea is that unconditional branches determine the global control flow of a program. Consequently, Shotgun uses a large part of its BTB to hold unconditional branches. Moreover,  Shotgun stores two footprints for each unconditional branch that show which instruction blocks are accessed around that branch and its target. Using these modifications, Shotgun offers better performance, mainly on workloads with larger instruction footprints.

        \section{Motivation}

We compare RDIP, Shotgun, and PIF to determine their advantages and disadvantages. We, then, discuss why we need to develop a new instruction prefetcher and how this prefetcher should be to address the shortcomings of the prior work. The details of the competing approaches, the simulated core, and the benchmarks are given in Section~\ref{sec:method}.

\begin{figure}[h]
    \centering
    \includegraphics[width=1.0\textwidth]{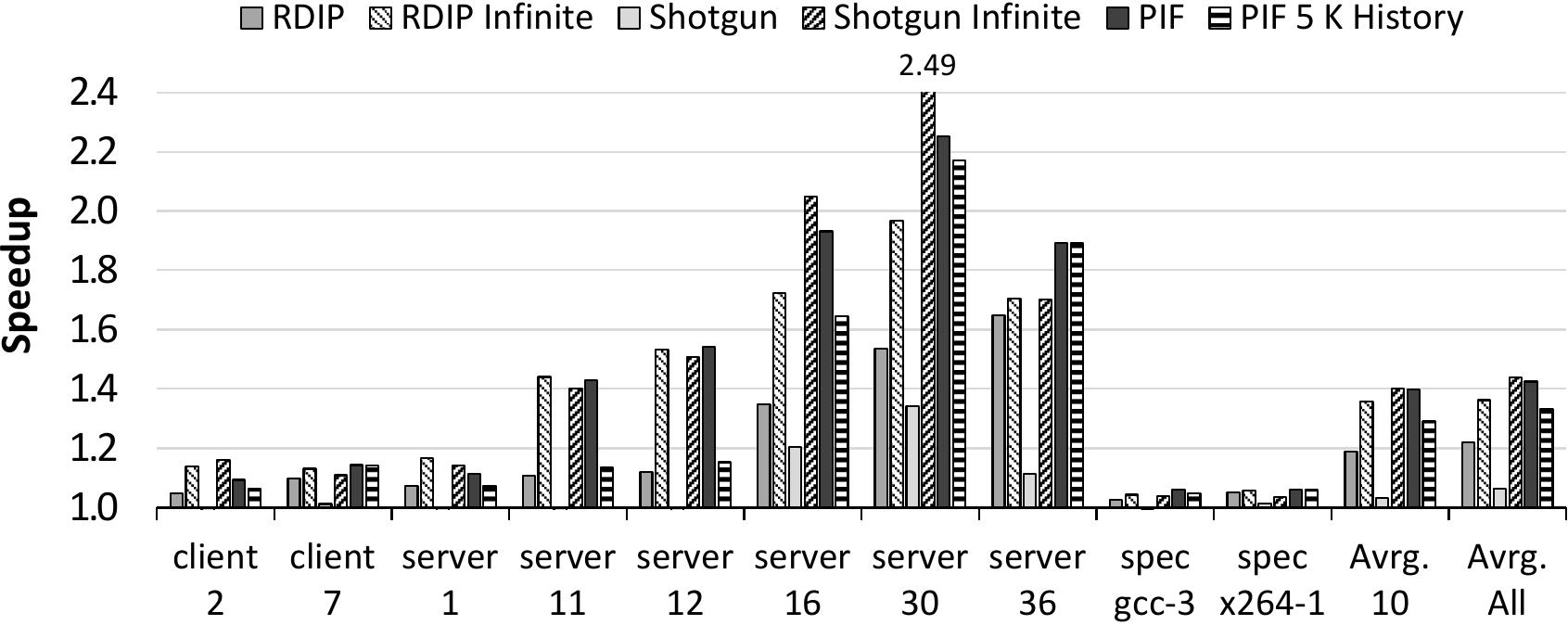}
    \caption{Speedup offered by state-of-the-art prefetchers. 
        \label{fig:speedup_motivation}}
\end{figure}

\subsection{Performance Comparison}

Figure~\ref{fig:speedup_motivation} compares the performance improvement of RDIP, Shotgun, and PIF over a baseline without a prefetcher. For RDIP and Shotgun, along with the authors-proposed configuration, we evaluate a configuration with infinite storage. Moreover, we evaluate an implementation of PIF in which the history buffer has 5~K entries while it has 32~K entries in the authors-proposed configuration. Results reveal three essential facts. First, PIF outperforms RDIP and Shotgun with a large gap. It means that the reduction in storage in RDIP and Shotgun is achieved at the cost of losing considerable speedup. Second, RDIP and Shotgun considerably fill this gap when infinite storage is available to them. As such, the performance gap in the original configuration is because RDIP and Shotgun are incapable of holding the large number of records that they require to prefetch effectively. Finally, PIF loses performance when the history buffer is reduced. It means that PIF needs a large history buffer to exploit the potential.

\subsection{Number of Prefetching Records}

To find out why RDIP and Shotgun suffer from metadata-miss, we should know what their prefetching records are and how they are stored. RDIP creates signatures that are the bit-wise XOR of the four top entries in the RAS. Moreover, it sets the last bit of signatures based on the type of control-flow change (return or call). The signatures are used to look up the Miss Table in which the addresses of missed blocks are recorded. As a result, RDIP should have a record for each observed signature in the Miss Table, and its number of required records is equal to the number of distinct signatures. We note that RDIP suggested a 4~K-entry Miss Table that is organized as a 4-way set-associative structure.   

On the other hand, Shotgun needs to store basic blocks in its BTBs. Shotgun discovers basic blocks one after another and inserts them into the FTQ and prefetches the blocks associated with those basic blocks. As a result, the prefetching record of Shotgun is a basic block, and BTB should be large enough to accommodate the basic blocks. To hold these basic blocks, Shotgun uses three BTBs that hold 2~K entries altogether. However, Shotgun attempts to prefill its BTBs to compensate for its relatively small BTBs. Nevertheless, Figure~\ref{fig:speedup_motivation} shows that even with the prefilling mechanism, the metadata-miss problem is still a considerable bottleneck.

Finally, PIF benefits from spatial regions. Each spatial region consists of a block address, called a \textit{trigger}, and a footprint that shows the bitmap of accessed blocks around the trigger. Using the footprint, PIF can detect the accessed blocks by keeping a single bit for each block at the cost of storing the full address of the trigger. PIF writes these spatial regions in a long circular history. Moreover, it uses an index table that records the latest entry of the history buffer in which a particular spatial region is recorded. PIF suggests an index table with 8~K entries that are organized as a 4-way set-associative structure and a 32~K-entry history buffer that is a circular buffer to hold the required prefetching records.

\begin{figure}[h]
    \centering
    \includegraphics[width=1.0\textwidth]{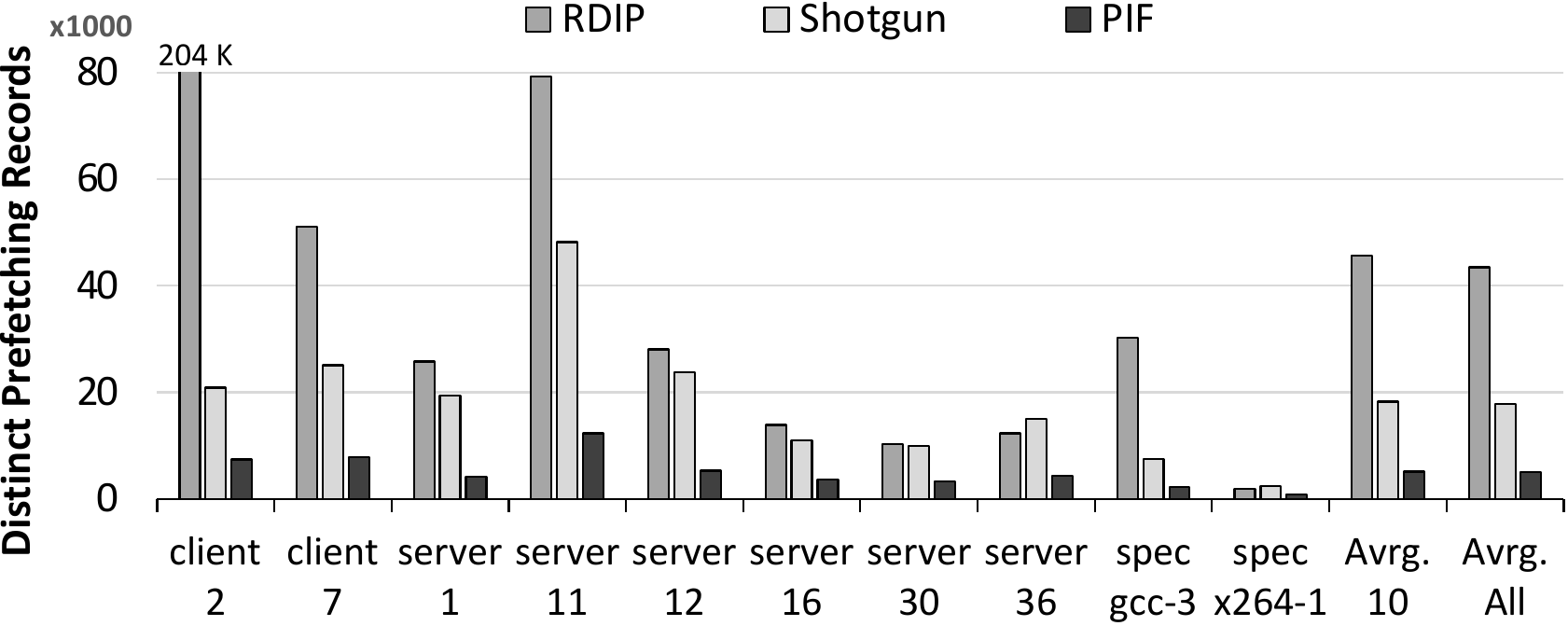}
    \caption{Number of distinct prefetching records.
        \label{fig:distinct_trigger}}
\end{figure} 

Figure~\ref{fig:distinct_trigger} shows the number of distinct records that are observed for each prefetcher. In other words, we count the number of signatures for RDIP, basic blocks for Shotgun, and spatial regions' trigger addresses for PIF. It can easily be inferred that RDIP and Shotgun have a significantly larger number of distinct prefetching records. Comparing these values to the number of entries that are suggested for RDIP's Miss Table and Shotgun's BTBs, we conclude that these approaches need orders of magnitude more entries to obtain the full potential. Moreover, we observe that PIF has a significantly smaller number of distinct records. The absolute value is close to 5~K on average, and an 8~K-entry index table can accommodate the records.

While Figure~\ref{fig:distinct_trigger} suggests that PIF has fewer distinct prefetching records, its design cannot exploit this advantage. Figure~\ref{fig:speedup_motivation} shows that by decreasing the number of history-buffer entries from 32~K to 5~K, the obtained speedup shrinks from 42.5\% to 32\%. This result corroborates a similar study in prior work~\cite{kaynak:shift}. The reason is that multiple instances of a spatial record may be written in PIF's history buffer. Consequently, while the number of distinct records is about 5~K, the history buffer should be much larger to hold all of the records successfully. Note that a version of PIF that has a 5 K-entry index table and a 5~K-entry history buffer requires 59 KB, which is still significant.

\subsection{Size of Prefetching Records}

Not only the number of distinct records but also the size of each record influences the storage overhead. In this section, we take a look at the records of various prefetchers. To make a fair and consistent comparison, we assume that prefetchers deal with a 46-bit address space.

\textbf{\textit{RDIP:}} Each entry in RDIP's Miss Table consists of a signature tag and three trigger addresses, each having an 8-bit footprint. As signatures are 32-bit long, and Miss Table is a 4~K-entry, 4-way set-associative structure, each signature tag is 22-bit long. On the other hand, each trigger address is 40 bits. Summing up altogether, the total storage cost of each Miss Table entry is 166 bits (over 20 bytes). 

\textbf{\textit{Shotgun:}} Shotgun associates its required information with the BTB and states that it is a metadata-free prefetcher. Nevertheless, to have a powerful Shotgun prefetcher, there is no other way than increasing the BTB size. Unfortunately, BTB is a storage-hungry component as it requires two instruction addresses, the branch address, and the branch target. While Shotgun proposes three separate BTBs and they have some differences, we consider U-BTB in this study, as it is the largest BTB of Shotgun. 

Considering a 2~K-entry BTB that is organized as a 4-way set-associative structure, the tag of the basic-block address needs 37 bits. The target address is 46 bits. Moreover, an entry needs 5 bits for the basic block's size and 1 bit for the branch type. Finally, Shotgun adds two Call and Return footprints to a BTB entry, and each footprint is 8 bits. Altogether, every BTB entry requires 105 bits (over 13 bytes).

\textbf{\textit{PIF:}} Each entry of the index table has a spatial-region tag and a pointer to a 32~K-entry history buffer. As the trigger address of each spatial region is a block address, a spatial-region tag is 29-bit long. Moreover, the pointer requires 15 bits to index a 32~K-entry history buffer. As a result, 44 bits should be used for an entry in the index table. Moreover, every entry of the history buffer is a spatial region. A spatial region needs 40 bits for the trigger address and 8 bits for the footprint. As a result, 48 bits are used for an entry in the history buffer. The sum of the number of bits of entries in the index table and history buffer is 92 bits (over 11 bytes).


\begin{figure}[h]
    \centering
    \includegraphics[width=1.0\textwidth]{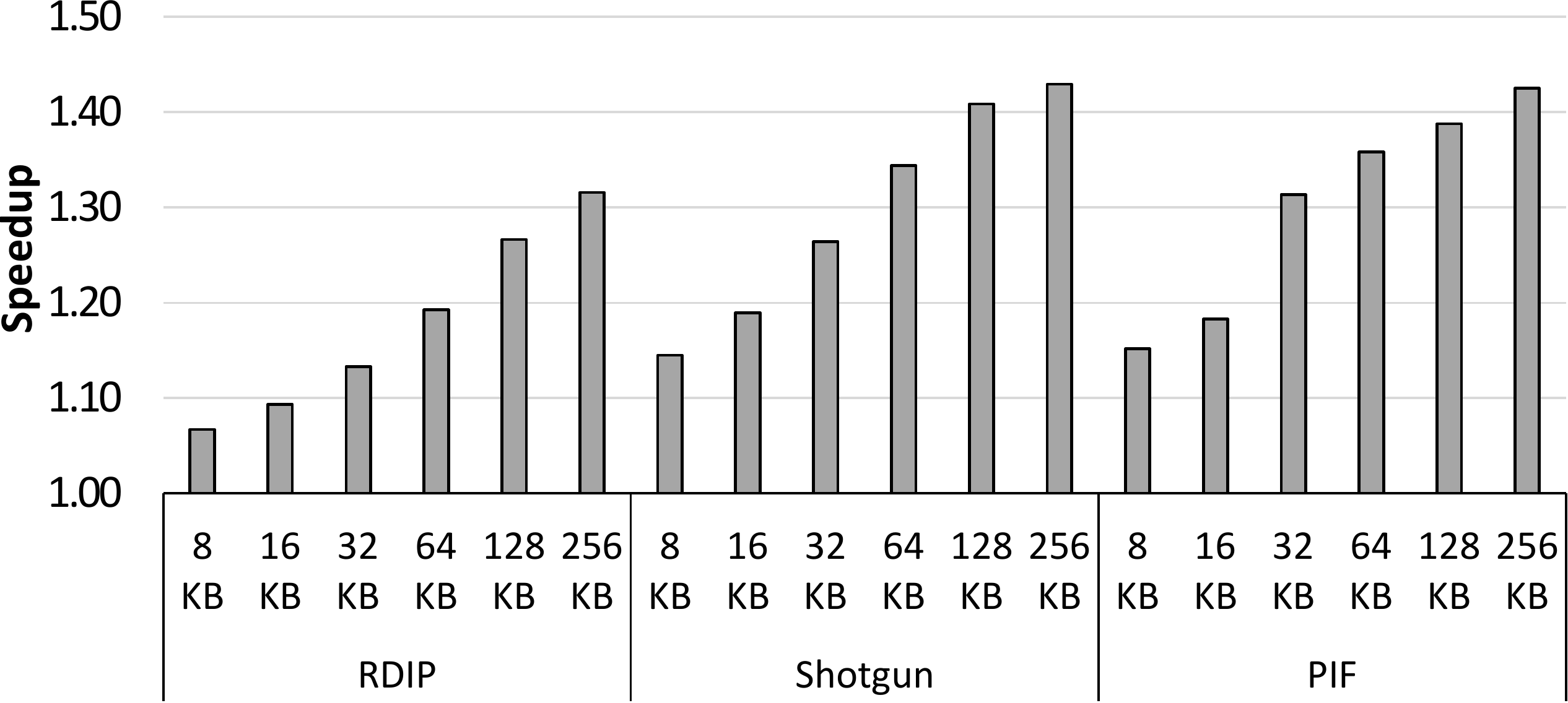}
    \caption{Speedup of RDIP, Shotgun, and PIF with an equal storage budget, averaged across 50 benchmarks.
        \label{fig:size_8_to_256}}
\end{figure}   

This study shows that due to having a large number of distinct records and the high storage cost of each record, prior prefetchers either do not offer the full potential or need an impractically large storage budget. Figure~\ref{fig:size_8_to_256} experimentally verifies this conclusion by showing the speedup of these prefetchers when we change the storage budget from 8 to 256~KB. The reported speedup is the average across all 50 benchmarks (See Section~\ref{sec:method}). We assume that the baseline design has a 2~K-entry BTB. In the case of Shotgun, we use the storage budget to enlarge the BTB.

Generally, Shotgun and PIF offer a very close performance improvement across different storage budgets. However, Figure~\ref{fig:size_8_to_256} shows that RDIP lags behind the other prefetchers in all studied storage budgets. Finally, comparing Figure~\ref{fig:size_8_to_256} and Figure~\ref{fig:speedup_motivation} reveals that even a 256~KB storage budget is not sufficient for RDIP to reach its full potential, which is offered by RDIP infinite.

\subsection{Spatial Regions, a Common Feature}
\label{sec:sr_cp}

All of the three prefetchers use a structure similar to PIF's spatial regions. RDIP's Miss Table and Shotgun's U-BTB both have footprints to encode the prefetch candidates associated with the trigger. Such a structure is used because accessed or missed blocks have high spatial correlations, and a footprint can encode the blocks in a lightweight manner. However, surprisingly, none of the prior work used a simple table to record spatial-region footprints while the table is looked up by the trigger address. 

RDIP, Shotgun, and PIF not only successfully detect the \textbf{current} spatial region but also have the ability to find the \textbf{successor} spatial regions. By providing this feature, the prefetcher can (a) prefetch the trigger address of spatial regions and (b) offer excellent timeliness. Providing this feature contributed to many of the complexities of these prefetchers. RDIP associates the current signature's misses to the prior signature. As a result, the prefetched misses are one signature (or equivalently, one call or return) ahead of the fetch stream. Shotgun follows basic blocks one after another to reach the successive U-BTB or RIB hit to prefetch the corresponding footprints. PIF writes the sequence of spatial regions in its history buffer, and consequently, the successive spatial regions are the subsequent entries in the history buffer. 


\subsection{What Should be Done?}

We know that the spatial region is an excellent prefetching record because of the high spatial correlation among accesses, the compactness of the footprint, and the fewer number of distinct spatial regions' trigger addresses as compared to other record types (See Figure~\ref{fig:distinct_trigger}). Consequently, a prefetcher that uses spatial regions as its prefetching records does not suffer from the two crucial limitations of prior work. Moreover, prior work suffered from the mechanism of identifying the subsequent spatial regions. Unfortunately, a spatial region by itself does not identify the next spatial region. In Section~\ref{sec:design_HOBP}, we discuss how this feature can be provided without the need for complicated and storage-hungry approaches used in the prior work.

        \section{MANA Prefetcher}
\label{sec:mana_prefetcher}

MANA creates the spatial regions using a spatial region creator and stores them in a set-associative table named MANA\_Table. Each spatial region is also associated with a pointer to another MANA\_Table entry in which its successor is recorded. A pointer is sufficient to benefit from temporal correlations, as prior temporal prefetchers showed that recently-accessed addresses tend to recur~\cite{ferdman:tifs, ferdman:pif, wenisch:tms, wenisch:stms, ferdman:phdthesis}. To reduce the storage cost, MANA exploits this observation that there are a small number of distinct high-order-bits patterns. This phenomenon is because the code base of a program has a high spatial locality and is much smaller than the size of the physical memory~\cite{bakhshalipour2019reducing, bakhshalipour2018stacked}. Consequently, instead of recording the complete trigger address that is the largest field of spatial regions, MANA uses the pointers to the observed high-order-bits patterns that need a considerably fewer number of bits. 

\subsection{Spatial Region Creator}

Spatial Region Creator (SRC) is responsible for creating MANA's prefetching records. Spatial regions consist of a trigger address and a footprint that shows which instruction blocks are observed in the neighborhood of the trigger address. SRC tracks the retire-order instruction stream and extracts its instruction blocks. If the current instruction block is different from the last observed instruction block, SRC attempts to assign this new instruction block to a spatial region. SRC has a queue of spatial regions named Spatial Region Queue (SRQ). After detecting a new instruction block, SRC compares this instruction block with SRQ entries. If this block falls in the address space covered by one of the SRQ entries, SRC sets the corresponding bit in that spatial-region footprint. Otherwise, SRC dequeues an item from SRQ, creates a new spatial region whose trigger address is the new instruction block, resets the footprint, and enqueues it in the SRQ.

\subsection{MANA\_Table}

When SRC dequeues an entry from SRQ to enqueue a new spatial region, the evicted spatial region is inserted into MANA\_Table. MANA\_Table stores the spatial regions and uses a set-associative structure with Least Recently Used (LRU) replacement policy that is looked up by the trigger address of the spatial region. Upon an insertion, if a spatial region hit occurs, the spatial region's footprint is updated with the latest footprint. Otherwise, the LRU entry is evicted, and the new spatial region is inserted into MANA\_Table. 

\subsection{Finding the Next Spatial Region}

We include in MANA\_Table's prefetching record a pointer to another entry of MANA\_Table to provide a sufficient lookahead to prefetch subsequent spatial regions. Whenever a spatial region is inserted into MANA\_Table, MANA records its location. By knowing this location, when MANA records a new entry in the table, the pointer of the last recorded spatial region is set to the location of the new entry. Using these pointers, MANA can chase the spatial regions one after another by iteratively going from a spatial region to its successor.

\subsection{Trigger Address is too Long!} 
\label{sec:design_HOBP} 

Considering a 46-bit address space and a 4~K-entry 4-way set-associative MANA\_Table, each record requires a 30-bit trigger-address tag, an 8-bit footprint, and a 12-bit pointer to the successor. The 8-bit footprint is derived from prior work~\cite{ferdman:pif, kaynak:shift, kumar:shotgun}. However, in Section~\ref{sec:spatial_region}, we choose the appropriate footprint type for MANA. To further reduce the storage cost, we observe that there is a considerable similarity between the high-order bits of the instruction blocks, and there are a few distinct patterns due to the high spatial locality of the code base of programs. As a result, we divide the trigger address tag into two separate parts, a partial tag, and the rest of the high-order bits. 

We store the partial tag in MANA\_Table and the rest of the bits in a separate structure. The division of tag should be done in a way to minimize the storage overhead. If we devote more bits for the partial tag, we will have fewer high-order-bits patterns (HOBPs), but we need to store longer partial tags in MANA\_Table. On the contrary, if we devote a fewer number of bits to the partial tag field, we will encounter more distinct HOBPs. In the evaluation section, we show how to divide the tag to minimize the overhead.

MANA stores HOBPs in a set-associative table named high-order-bits patterns' table (HOBPT). Every new observed HOBP is inserted into HOBPT. Moreover, each MANA\_Table record has a \textit{HOBP index}, which points to a HOBPT entry in which the corresponding HOBP is recorded. 

\begin{figure}[h]    
    \centering
    \includegraphics[width=1.0\textwidth]{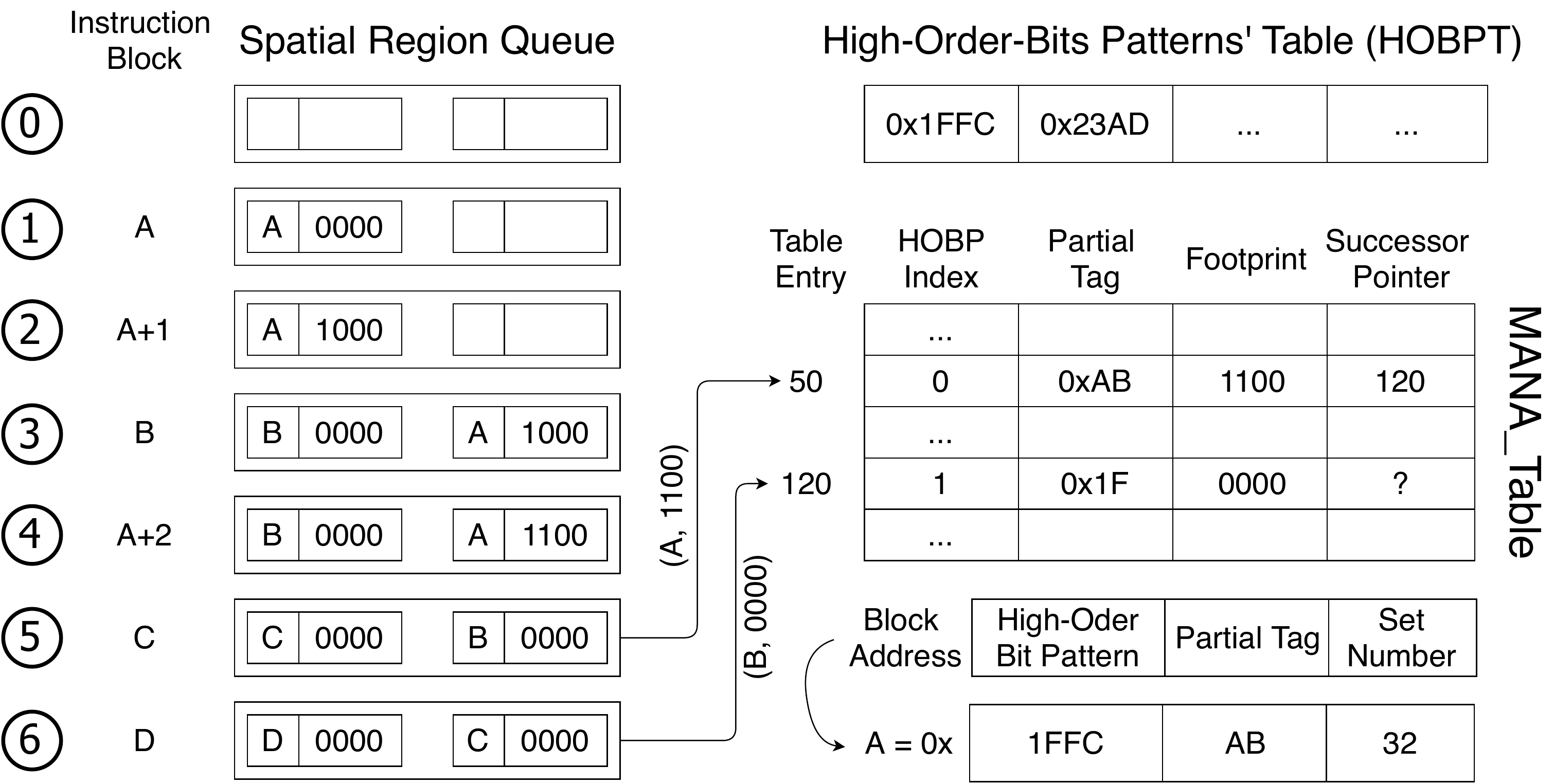}
    \caption{Overview of MANA recording spatial regions.
        \label{fig:design}}
\end{figure}

\subsection{Example}

Figure~\ref{fig:design} shows how MANA records the spatial regions. The configuration of MANA is for illustrative purposes. The actual configuration is determined through sensitivity analyses in Section~\ref{sec:eval}. The shown process has six steps. SRQ holds two spatial regions whose footprints contain four instruction blocks ahead of the trigger address. The figure also shows the state of HOBPT and MANA\_Table. 

We represent a spatial region as '(A, XXXX)' where 'A' is the trigger address, and 'XXXX' is the footprint of the following four blocks.  In the beginning, SRQ is empty. SRC tracks the block address of the retired instructions. In the first step, the retired instruction belongs to block \textit{A}. SRC looks up SRQ, and as it is empty, SRC finds no match. As a result, SRC creates a new spatial region which is '(A, 0000)'. In Step 2, the retired instruction is from block \textit{A+1}. SRC looks up SRQ and finds out that \textit{A+1} falls in the address space covered by '(A, 0000)' and sets the corresponding bit in its footprint. In the next step, the observed instruction block is \textit{B} that does not fall in any of the already created spatial regions. Consequently, SRC creates a new spatial region and enqueues it in the SRQ. In Step 4, similar to Step 2, another bit is set in the spatial region's footprint, whose trigger address is \textit{A}. In Step 5, SRC cannot match the instruction block with any of the SRQ entries. Moreover, SRQ is full. As a result, it dequeues an entry from SRQ to provide a space for the new spatial region. SRC inserts '(C, 0000)'  into SRQ. Moreover, the evicted spatial region is sent to MANA\_Table. Similarly, in Step 6, '(D, 0000)' is enqueued in SRQ, and '(B, 0000)' is evicted and stored in MANA\_Table.

We now describe how spatial regions are placed in MANA\_Table. For simplicity, we consider a direct-mapped MANA\_Table with 256 entries. Suppose that the address of instruction block \textit{A} is 0x1FFCAB32. Due to MANA\_Table's structure, the eight lower-order bits of \textit{A} are used to determine the table index in which this spatial region should be stored. Because the lower-order bits are 0x32 or 50 in decimal number representation, as the figure shows, this spatial region is inserted into the 50th entry of the table. MANA\_Table exploits the commonality among the high-order bits of the instruction block addresses. This means that MANA compares the 16 high-order bits of \textit{A} with the observed patterns in HOBPT. As MANA finds a match in the first entry of HOBPT, zero is recorded in the \textit{HOBP index} field of MANA\_Table. This way, instead of keeping 16 bits for higher-order bits, we need only four bits assuming that HOBPT keeps 16 distinct patterns. Moreover, we use an 8-bit partial tag to store the remaining bits of the trigger address tag that are not stored in HOBPT. Figure~\ref{fig:design} shows how an instruction cache block address can be constructed by combining the HOBP index, the partial tag, and the set number in which a record is stored.

MANA also updates the successor pointer. When MANA inserts '(A, 1100)' into MANA\_Table, it also stores the table index in which this record is inserted (i.e., 50). Later on, when MANA inserts the next record '(B, 0000)' in the 120th row of MANA\_Table, it also goes to row 50 and sets the successor pointer to 120. It means that when MANA prefetches the instruction blocks recorded in '(A, 1100)', it can easily prefetch '(B, 0000)' by following the pointer. After setting the successor pointer, MANA updates its last inserted record from 50 to 120. 

\subsection{Replaying}
\label{sec:replaying}

MANA takes advantage of the stream address buffer (SAB), which is previously used by prior temporal prefetchers~\cite{ferdman:pif, kaynak:shift, kaynak:confluence}, to prefetch for the L1-I cache. SAB is a fixed-length sequence of spatial regions that are created by chasing the spatial regions one after another from the pointers that are stored in MANA\_Table. Moreover, SAB has a pointer to the MANA\_Table entry that the last spatial region is fetched from and inserted into SAB. 


SAB has three goals. First, it enables MANA to go sufficiently ahead of the retire-order instruction stream to issue timely prefetches. Second, SAB helps MANA to know the instruction blocks that are already prefetched and the current lookahead of them. Finally, by tracking the spatial regions that are already prefetched, SAB enables MANA to eliminate redundant and repetitive prefetch requests.

MANA attempts to have a fixed lookahead ahead of the fetch stream to prefetch the trigger address of the successor spatial regions and also to have timeliness. This lookahead is defined as the number of spatial regions that MANA prefetches ahead when it observes an instruction cache block. MANA tracks the fetch stream and extracts its instruction block addresses. If the block address falls in the address space that is covered by a spatial region in a SAB, MANA checks the number of spatial regions that are prefetched after the matched spatial region, and hence, are inserted into SAB. If this number is lower than the lookahead, MANA chases the spatial regions using SAB's pointer to MANA\_Table to have sufficient lookahead. If MANA finds no SAB associated with the block address, it considers the instruction block as the trigger address of a spatial region and looks up MANA\_Table to find the corresponding spatial region. If MANA\_Table finds a match, MANA evicts the LRU SAB entry (if it has multiple SABs) and creates a new SAB by inserting the found spatial regions into SAB and chasing its successor pointer to find the next spatial region. MANA repeats this process until the number of inserted spatial regions into SAB reaches the predefined lookahead depth. Finally, MANA extracts the instruction blocks that are encoded in the footprint of the inserted spatial regions and prefetches them.

        \section{Methodology}
\label{sec:method}

\subsection{Simulation Framework}

To evaluate our proposal, we use ChampSim simulator ~\cite{champsim_github} with the configurations shown in Table~\ref{table:method}. We made substantial changes to ChampSim to accurately model the frontend of a processor. Among them, we model BTB in ChampSim, which is not modeled in the baseline implementation. Moreover, we simulate BTB miss and branch direction/target misprediction stalls in the address generation component as a result of modeling BTB\footnote{The baseline implementation only models branch direction misprediction stalls.}. Furthermore, we model wrong-path fetches when a taken BTB miss or a branch direction/target misprediction happens. Modeling the wrong-path fetches is important as they may pollute the cache hierarchy if they eventually become useless. Or they may become useful as observed in~\cite{kumar:boomerang}, and hence, lower the usefulness of a prefetcher. The evaluated prefetchers are triggered on instruction or block addresses touched in the wrong-path, which affects their behavior. We find these changes are crucial to have a fair and accurate evaluation of the competing prefetchers in the context of ChampSim.

\subsection{Benchmarks}

We use public benchmarks that are provided by the first instruction prefetching championship (IPC-1)~\cite{IPC_1}. This package contains 50 benchmarks, including eight for the clients' execution, as well as 35 for servers, and seven from SPEC. While we execute all 50 benchmarks, as it is not possible to show all the results, we selected ten benchmarks that represent various observed behaviors. Moreover, we report the average of the ten selected benchmarks as well as the average of all 50 benchmarks. 

Each benchmark is executed for 50 million instructions to warm up the system, including the caches, the branch predictor, and prefetchers' metadata. The rest of the instructions (i.e., 50 million instructions) are used to collect the evaluation metrics, including Instruction Per Cycle (IPC).

\begin{table}[h]
    \sffamily
    \begin{center}
        \caption{Evaluation Parameters}
        \label{table:method}
        {
            \resizebox{1.0\textwidth}{!}{
                \begin{tabular}{| l | l |}
                    \hline
                    {\bf Parameter}                         & {\bf Value}  \\
                    \hline
                    \hline
                    \multirow{2}{*}{Core}       & 14~nm, a single 4~GHz OoO core \\
                    & 352-entry ROB, 128-entry Load Queue, 72-entry Store Queue \\
                    \hline
                    \multirow{2}{*}{Fetch Unit} & 32~KB, 8-way, 64B block size, 4-cycle latency \\ & 
                    hashed-perceptron branch predictor~\cite{tarjan2005merging}, 2K-entry Branch Target Buffer, 8-entry MSHRs \\
                    \hline
                    \mbox{L1-D} Cache & 48~KB, 12-way, 64B block, 5-cycle latency, 16-entry MSHRs, next-line prefetcher  \\
                    \hline
                    L2 Cache & 512~KB, 8-way, 10-cycle latency, Signature Path Pattern (SPP)~\cite{kim2016path} prefetcher \\
                    \hline
                    \multirow{1}{*}{Last Level Cache}        & 2~MB, 16-way, 20-cycle latency, 64-entry MSHRs \\
                    \hline
                \end{tabular}
            }
        } 
    \end{center}
\end{table}

\subsection{Competing Proposals}

\textbf{\textit{baseline:}} Table\mbox{~\ref{table:method}} summarizes our baseline core's configuration. All performance improvements are measured against this baseline. 

\textbf{\textit{RDIP:}} RDIP~\cite{kolli2013rdip} uses the top four entries of RAS to create the signatures. Moreover, we model a 4~K-entry, 4-way set-associative Miss Table, where each entry holds three trigger addresses and the associated footprint of the observed instruction cache misses. All the parameters are chosen based on the original proposal~\cite{kolli2013rdip}. This prefetcher imposes 83 KB storage to the baseline design with no instruction prefetcher.

\textit{\textbf{Shotgun:}} We model Shotgun~\cite{kumar:shotgun} with a 1.5~K-entry U-BTB, a 128-entry C-BTB, and a 512-entry RIB, as suggested in the original proposal. Moreover, Shotgun uses a 64-entry instruction cache prefetch buffer and a 32-entry BTB prefetch buffer. Shotgun imposes 6~KB storage cost to the baseline design; 4 KB comes from the prefetch buffers, and the rest is because of the changes made to the BTB.

\textbf{\textit{PIF:}} PIF~\cite{ferdman:pif} records the sequence of spatial regions in a circular history buffer of 32~K spatial regions. To find a record in the history buffer, PIF uses an index table that holds pointers to history buffer's entries. We model a 4-way set-associative index table with 2 K sets, as in the original proposal. PIF imposes over 236 KB of storage overhead to the baseline design. As in the original proposal, the temporal compactor contains eighteen spatial regions, the lookahead is five, and four SABs are used where each one tracks seven consecutive spatial regions~\cite{ferdman:pif, kaynak:shift}.

\textbf{\textit{MANA:}} MANA stores the spatial regions in a 4~K-entry table of 1~K sets. Each MANA\_Table record consists of 7 bits to indicate the HOBP index, a 2-bit partial tag, an 8-bit footprint, and a 12-bit pointer to the successor spatial region. Moreover, it uses a 128-entry, 8-way set-associative HOBPT. MANA needs a 15 KB storage budget for its metadata. 

\textbf{\textit{Minimum Latency (MinLat) L1-I:}} MinLat L1-I is used to show the potential of instruction prefetching. In MinLat L1-I, lower levels of the caching hierarchy spend a single cycle when they serve instruction blocks' requests.


\textbf{\textit{MinLat L1-I + Perfect BTB:}} In this implementation, an ideal BTB is also used along with the MinLat L1-I that only faces compulsory misses. 

\subsection{BTB Prefilling} 

In recent proposals, the instruction prefetching problem is twisted to BTB prefilling problem to offer a unified solution to the frontend bottleneck~\cite{kaynak:confluence, kumar:boomerang, kumar:shotgun}. The main idea is that an instruction block has the required information to fill in the missing BTB entries. A simple instruction pre-decoder decodes the fetched and prefetched blocks to extract the branches. Then, these branches are inserted into the BTB to avoid BTB misses. In our evaluation, we assume that RDIP, PIF, and MANA also benefit from this BTB prefilling mechanism. It also provides a more fair comparison with Shotgun as it benefits from BTB prefilling mechanism in its design. However, in section~\ref{sec:instruction_prefetching_abilities}, we evaluate competing proposals only based on their instruction prefetching abilities.

        \section{Evaluation Results}
\label{sec:eval}

\subsection{Parameter Selection}

We start with an initial MANA prefetcher and run sensitivity analyses to tune the parameters. In the initial MANA, the lookahead is five, SRQ size is eighteen, there are four SABs, each having twelve entries. Prior work has shown that this configuration successfully exploits the potential~\cite{ferdman:pif, kaynak:shift}.
Moreover, MANA\_Table holds 4~K entries in a 4-way set-associative structure. We start with a 4~K-entry table because we found MANA creates less than 4~K distinct records, on average\footnote{In Figure~\ref{fig:distinct_trigger}, we show that PIF creates 5~K records, on average. MANA creates a smaller number of distinct records because instead of a decoupled spatial and temporal compactors used in PIF, it uses a coupled approach in its SRC that helps it to more efficiently create its spatial regions.}. As a result, we expect that such a table can hold the required prefetching records.

\subsubsection{Spatial Region Type}
\label{sec:spatial_region}

\begin{figure}[h]
    \centering
    \includegraphics[width=1.0\textwidth]{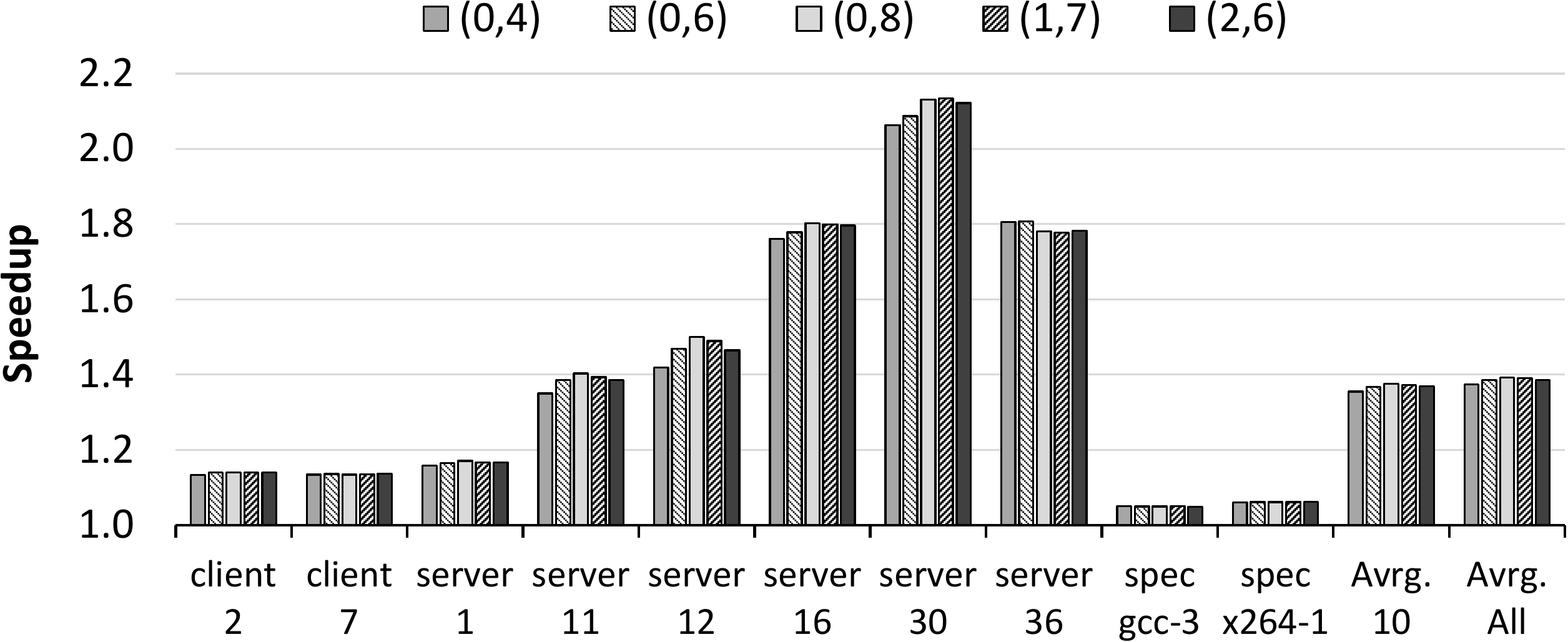}
    \caption{Speedup of various region types.
        \label{fig:region_type}}
\end{figure}

The spatial region type determines the length of the spatial region's footprint and the instruction cache blocks that are encoded into it. We use (X, Y) notation to represent a spatial region that holds X blocks behind and Y blocks ahead of the trigger block. Such a spatial region holds X+Y+1 instruction blocks. Note that the trigger block is held implicitly and does not need to have a dedicated bit in the footprint. Some pieces of prior work have used (2, 6) spatial regions~\cite{ferdman:pif, kolli2013rdip, kumar:shotgun}. However, we examine different spatial regions to find the best performing one. Figure~\ref{fig:region_type} shows MANA's speedup when (0, 4), (0, 6), (0, 8), (1, 7), and (2, 6) spatial regions are used. Results show that the gap between these regions is negligible. However, (0, 8) performs slightly better than the other types. As such, we use this region type in the rest of the evaluation. We note that (0, 4) can be a good design choice as well. It offers competitive speedup and requires four fewer bits, which provides a storage-saving opportunity. 

\subsubsection{SRQ Length}

\begin{figure}[h]
    \centering
    \includegraphics[width=1.0\textwidth]{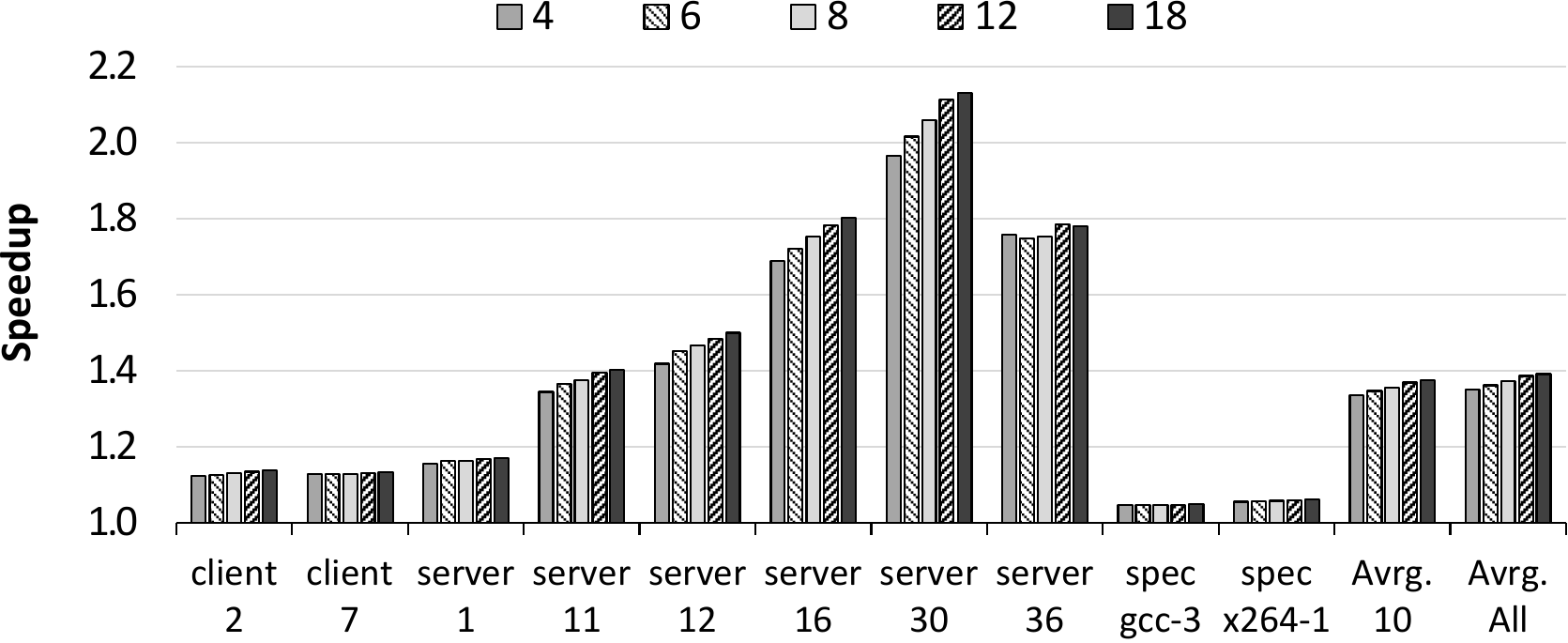}
    \caption{Speedup of various SRQ sizes.
        \label{fig:SRQ_length}}
\end{figure}

The next parameter that we are going to set is SRQ length. A longer SRQ holds the spatial regions for a more extended period. It enables SRC to better associate the observed instruction blocks to the already created spatial regions in SRQ. However, a longer SRQ imposes more search overhead. Figure~\ref{fig:SRQ_length} shows MANA's speedup when we change SRQ length from four to eighteen. As expected, increasing SRQ length improves speedup. However, we choose an 8-entry SRQ to have a more practical and simpler search process.

\subsubsection{Prefetching Lookahead}
\label{sec:eval_lookahead}

\begin{figure}[h]
    \centering
    \includegraphics[width=1.0\textwidth]{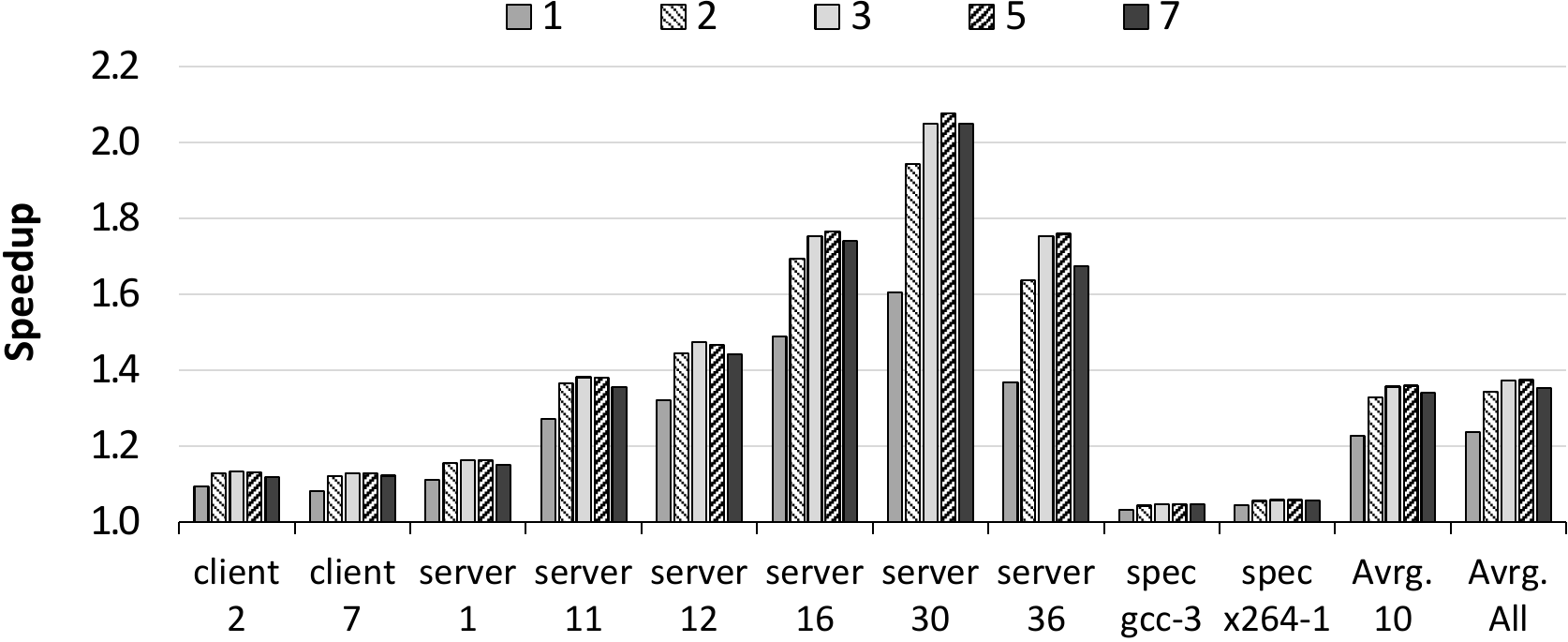}
    \caption{Effect of lookahead on speedup of MANA.
        \label{fig:speedup_lookahead}}
\end{figure}

We analyze the effect of various lookaheads on the MANA prefetcher. Figure~\ref{fig:speedup_lookahead} shows the obtained speedup when the lookahead is 1, 2, 3, 5, and 7. The results show that the lookahead of one lags significantly behind the others. Consequently, the pointer to the successor spatial region is necessary to offer good speedup. Moreover,  changing the lookahead from two to seven makes small differences. 

While the lookahead of one only offers 24\% speedup, comparing Figure~\ref{fig:speedup_lookahead} and Figure~\ref{fig:speedup_motivation} reveals that even MANA with the lookahead of one outperforms RDIP and Shotgun. When lookahead is one, MANA does not need the ability to chase the spatial regions. As a result, we can remove the pointer to the successor region field from MANA\_Table to reduce the storage overhead. Moreover, it does not need a full tag because MANA only prefetches the instruction blocks that are in the proximity of the trigger address. MANA needs the HOBP index, the partial tag, and the set number to create the trigger address of the spatial region when it chases the pointers. However, when the lookahead is one, MANA looks up MANA\_Table using a trigger address that is completely known. Just to determine that an entry is in the table or not, MANA needs a partial tag. We find that the 8-bit partial tag is sufficient to separate MANA\_Table hits from misses. This way, each MANA\_Table record contains an 8-bit partial tag and an 8-bit footprint, eliminating the need for a HOBP index. Considering a 4~K-entry MANA\_Table, 24\% speedup can be achieved using only 8 KB of storage.  

\begin{figure}[h]
    \centering
    \includegraphics[width=1.0\textwidth]{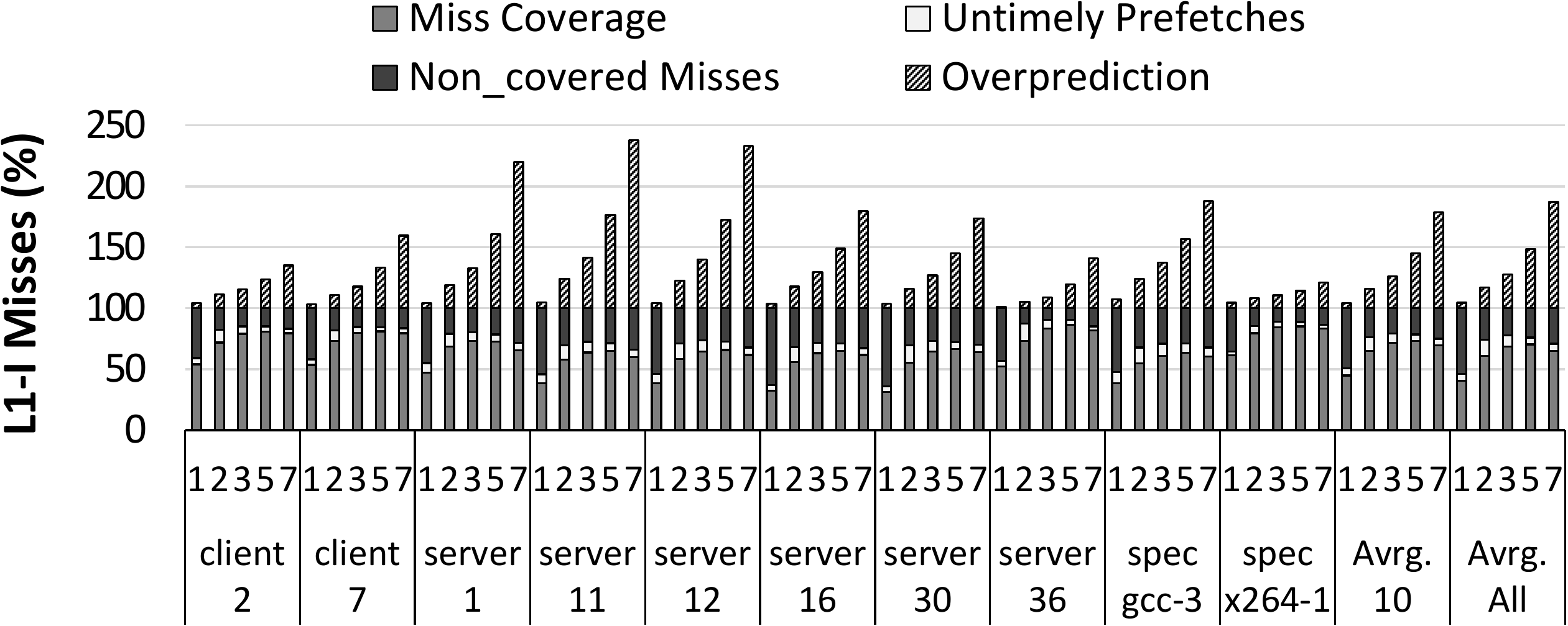}
    \caption{Effect of lookahead on MANA miss coverage.
        \label{fig:coverage_lookahead}}
\end{figure}

We also evaluate the coverage and overprediction of MANA when its lookahead varies from one to seven. A larger lookahead may offer better speedup but may cause lots of useless prefetches because of going too far ahead of the execution stream. Figure~\ref{fig:coverage_lookahead} clearly shows this effect. In this figure, miss coverage, non\_covered misses, untimely prefetches, and overprediction are shown. Miss coverage shows the fraction of misses that are eliminated by a prefetcher. The encountered misses are divided into two separate categories. The first type includes those misses that MANA did not send a prefetch for, and the second type contains cache misses that a prefetch has been sent for, but the demand arrived before the prefetch reply. The overprediction is the number of useless prefetches to the number of baseline misses and is used to evaluate prefetchers' accuracy~\cite{golshan2020harnessing}.

Figure~\ref{fig:coverage_lookahead} shows that increasing the lookahead from one to two improves the miss-coverage from 40\% to 61\%. The considerable better miss coverage is because when lookahead is one, MANA is not able to prefetch the trigger address blocks, and it can only prefetch the footprint. However, when the lookahead is two, MANA can also prefetch the trigger address of the successor spatial regions.
However, by going from one to two, the untimely prefetches have been increased from 5\% to 13\%. It means that MANA issues more prefetch requests, but they are not sufficiently timely to completely hide the fetch access latency. Going from two to three mitigates this problem: the untimely prefetch requests are decreased from 13\% to 9\%, and the miss-coverage reaches 69\%. Increasing the lookahead from three to five and seven only makes negligible differences to the miss-coverage and the untimely prefetches. However, the overprediction increases significantly. Based on Figure~\ref{fig:speedup_lookahead} and~\ref{fig:coverage_lookahead}, we set MANA's lookahead to three in the rest of the evaluation as it offers a right balance between the obtained speedup and the overprediction. 

\subsubsection{SAB Count and Length}

In section~\ref{sec:replaying}, we described how MANA uses SAB to issue or filter the prefetch requests. A single stream whose length is at least equal to the lookahead is the smallest possible SAB configuration. However, tracking multiple SABs or a longer SAB may be helpful because of successfully capturing the recently prefetched spatial regions. Prior work has suggested using four SABs that each one tracks seven~\cite{ferdman:pif} or twelve~\cite{kaynak:shift} spatial regions. However, we find a negligible difference between different configurations both in terms of performance improvement and the ability to filter redundant prefetches. We use a single SAB tracking five spatial regions to have a simple and practical design.

\subsubsection{MANA\_Table Size}

\begin{figure}[h]
    \centering
    \includegraphics[width=1.0\textwidth]{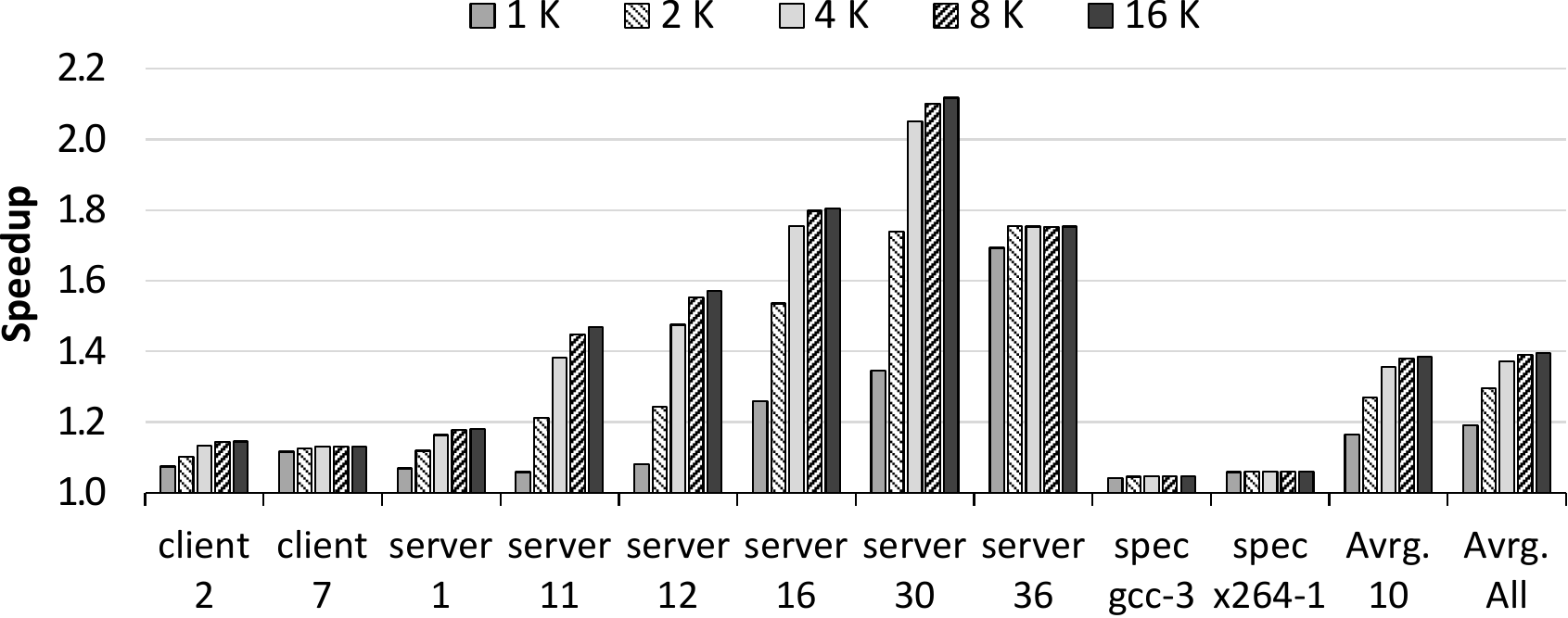}
    \caption{Speedup of various MANA\_Table sizes.
        \label{fig:speedup_tablesize}}
\end{figure}

Figure~\ref{fig:speedup_tablesize} shows the speedup of MANA for various MANA\_Table sizes. In prior experiments, MANA\_Table had 4~K entries in a 4-way set-associative structure. In this part, we still use a 4-way set-associative table but vary the number of entries from 1~K to 16~K. Figure~\ref{fig:speedup_tablesize} shows that a 4~K-entry table offers considerably better results as compared to 1~K- and 2~K-entry tables. Moreover, increasing the table size to 8~K and 16~K only offers a small improvement. Consequently, we set the number of MANA\_Table entries in the rest of the evaluation to 4~K.

\subsubsection{MANA\_Table Associativity}

\begin{figure}[h]
    \centering
    \includegraphics[width=1.0\textwidth]{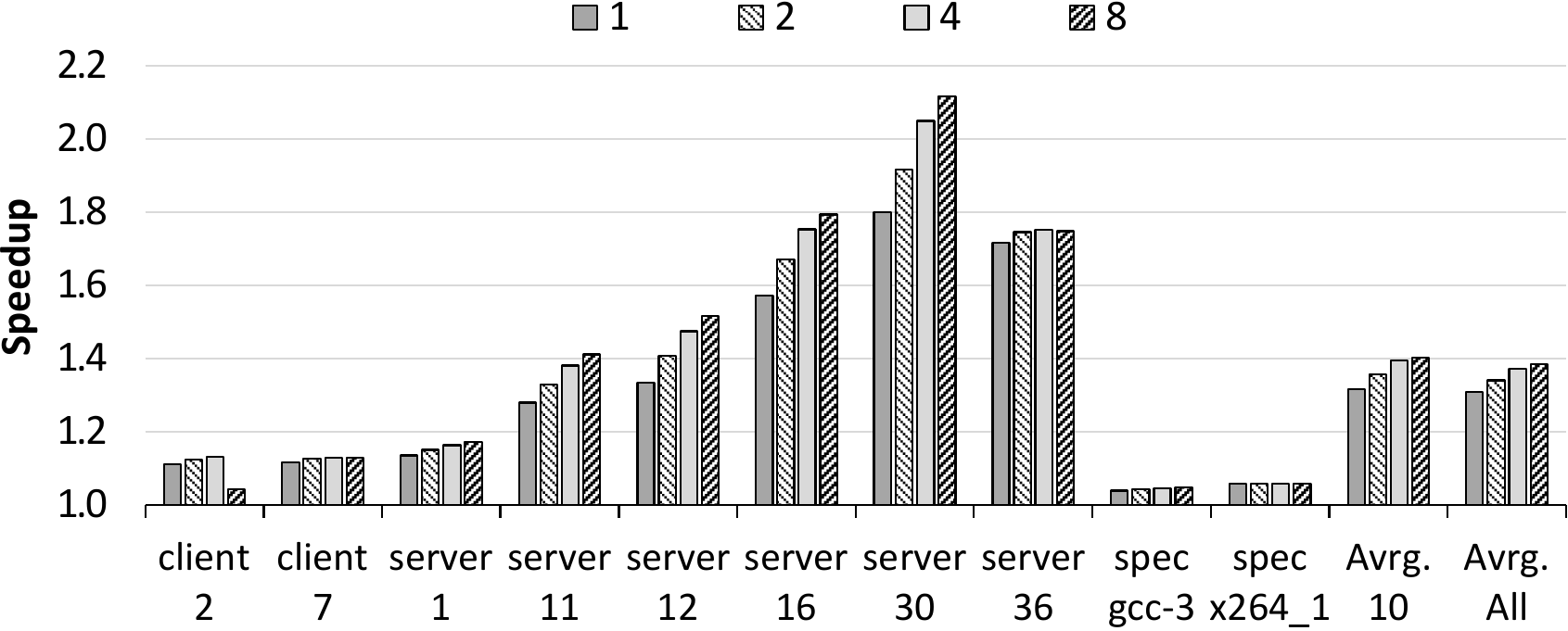}
    \caption{Effect of MANA\_Table's associativity on speedup
        \label{fig:associativity_speedup}}
\end{figure}

Figure~\ref{fig:associativity_speedup} shows how the speedup of MANA varies when we change MANA\_Table's associativity from 1 to 8. Increasing the associativity improves performance, however, the improvement from 4 to 8 is not considerable. As such, we use a 4-way set-associative structure for MANA\_Table.

\subsubsection{High-Order-Bits Patterns}

In Section~\ref{sec:design_HOBP}, we described how the storage requirements of MANA could be reduced by using a partial tag and the commonality of HOBPs. Table~\ref{table:HOBP} shows how changing the partial tag's length affects MANA's storage requirements. In this study, we assume that HOBPT is large enough to accommodate all observed patterns. We can infer that MANA requires lower storage when the partial tag's length is two. In this configuration, HOBPT needs to store up to 128 distinct HOBPs, which needs 0.44~KB.
Moreover, every MANA\_Table entry contains a 7-bit HOBP index to HOBPT. Altogether, HOBPT and MANA\_Table will require 14.94 KB storage. 

\begin{table}[h]
    \sffamily
    \begin{center}
        \caption{Effect of partial-tag length on MANA's storage requirements.}
        \label{table:HOBP}
        {
            \resizebox{1.0\textwidth}{!}{
            \begin{tabular}{| c | c | c | c | c |}
                \hline
                Partial Tag Bits & HOBP Index Bits & HOBPT Storage & MANA\_Table Storage & Sum \\
                \hline
                0 & 9 & 1.88 KB & 14.5 KB & 16.38 KB \\                
                \hline
                1 & 8 & 0.9 KB & 14.5 KB & 15.4 KB \\                
                \hline
                2 & 7 & 0.44 KB & 14.5 KB & 14.94 KB \\
                \hline
                5 & 5 & 0.1 KB & 15 KB & 15.1 KB \\
                \hline
                8 & 3 & 0.02 KB & 15.5 KB & 15.52 KB \\
                \hline
                11 & 3 & 0.02 KB & 17 KB & 17.02 KB \\
                \hline
            \end{tabular}
            }
        } 
    \end{center}
\end{table}

\subsection{Comparison with Prior Work}

In this section, we compare MANA prefetcher against state-of-the-art proposals.

\subsubsection{Performance Improvement}
\label{sec:performance_improvement}

\begin{figure}[h]
    \centering
    \includegraphics[width=1.0\textwidth]{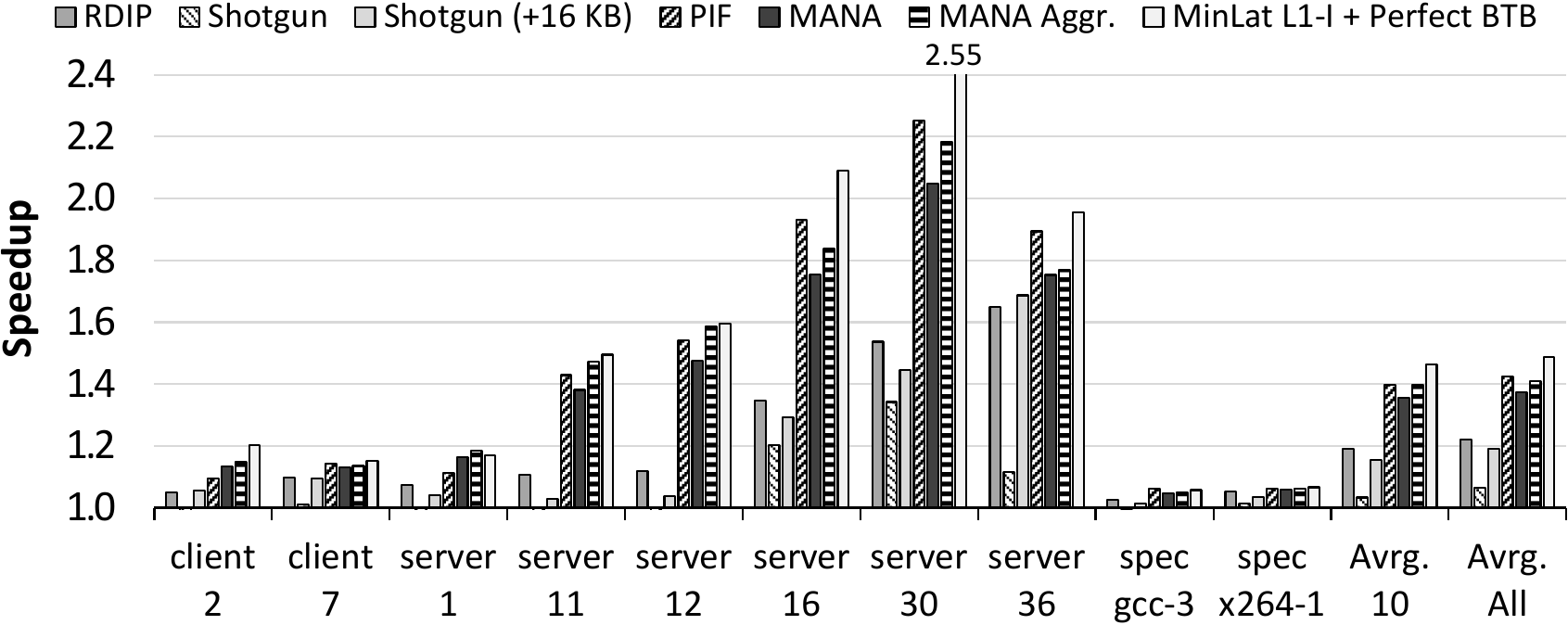}
    \caption{Speedup of competing prefetchers.
        \label{fig:speedup_state_of_the_arts}}
\end{figure}

Figure~\ref{fig:speedup_state_of_the_arts} shows the speedup of competing prefetchers. This figure reveals that MANA significantly outperforms RDIP and Shotgun and has only a small gap with PIF. RDIP, Shotgun, PIF, and MANA provide 22, 6.5, 42, and 38\% speedup on top of a baseline design without any prefetcher. It clearly shows how MANA outperforms its competitors with a large gap or offer competitive performance improvement with significantly smaller storage. Comparing MANA with \textit{MinLat L1-I + Perfect BTB}, we conclude that MANA offers 92\% of the performance that can be obtained by \textit{MinLat L1-I + Perfect BTB}.

To provide a fair comparison, in Figure~\ref{fig:speedup_state_of_the_arts}, we also evaluate two other prefetchers: \textit{Shogtun (+16 KB)} and \textit{MANA Aggressive}. Shotgun's BTB structures require 23.7 KB storage~\cite{kumar:shotgun}. As MANA imposes 15 KB storage overhead to the baseline design, we enlarge Shotgun's BTBs to 40~KB to provide the same storage to Shotgun as MANA's. This design is shown as \textit{Shotgun (+16 KB)}. Despite increasing the size of BTBs, \textit{Shotgun (+16 KB)} still lags behind MANA. Moreover, to have a fair comparison with PIF, we also evaluate an aggressive implementation of MANA shown as (\textit{MANA Aggr.}) in which, similar to PIF, the lookahead is five, the SRQ length is eighteen, the number of SABs is four, and each SAB tracks seven consecutive spatial regions. Besides, as PIF's index table holds 8~K pointer to the history buffer, we enlarge MANA\_Table to have an equal number of entries. This configuration needs 30~KB storage. Figure~\ref{fig:speedup_state_of_the_arts} shows that such an implementation has almost no gap with PIF. It means that the gap between MANA and PIF is due to our policy to set practical design parameters, and the aggressive implementation offers the level of performance that PIF offers while requires $7.8\times$ less storage.

\subsubsection{Miss Coverage}

\begin{figure}[h]
    \centering
    \includegraphics[width=1.0\textwidth]{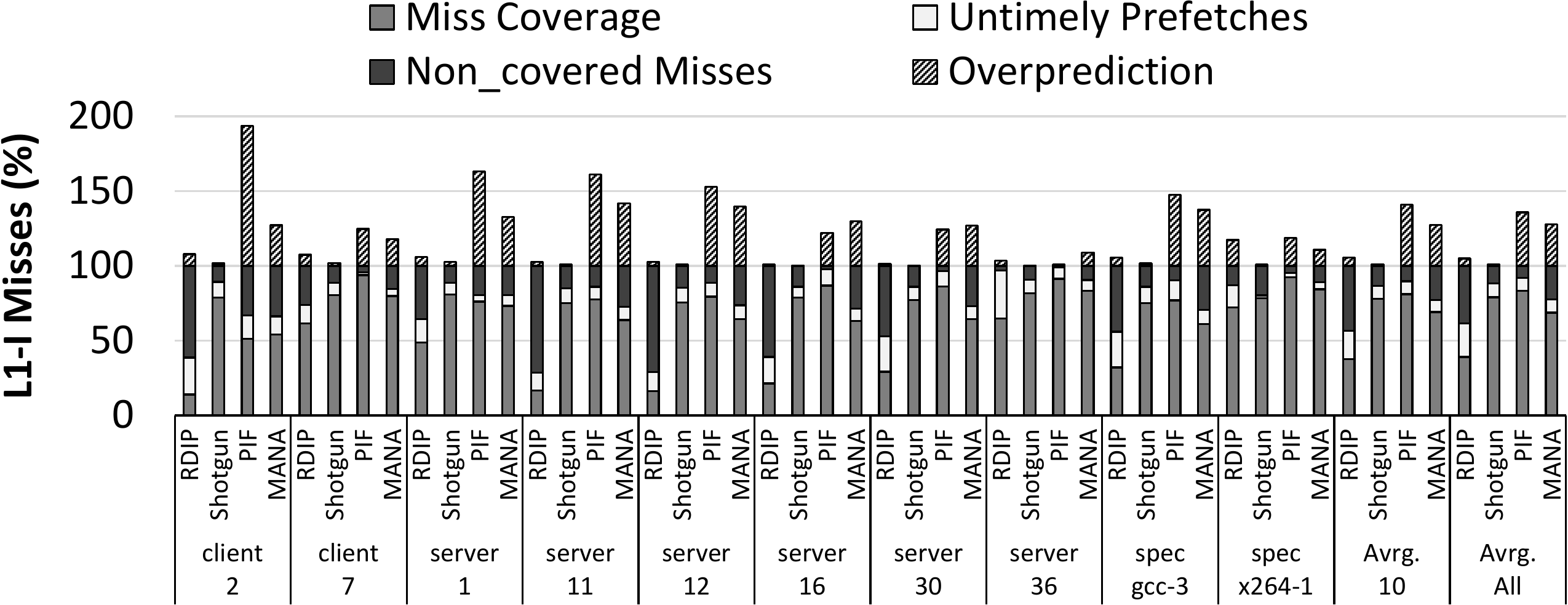}
    \caption{Miss coverage of competing prefetchers.
        \label{fig:coverage_state_of_the_arts}}
\end{figure}

Figure~\ref{fig:coverage_state_of_the_arts} shows how competing prefetchers perform in terms of covering the L1-I cache misses. RDIP covers a smaller fraction of misses and also has significant untimely prefetches that result in lower performance improvement. On the other hand, we see a considerably large miss coverage from Shotgun despite its poor performance improvement. The reason is that Shotgun stalls feeding the fetch engine when a BTB miss occurs. In such cases, Shotgun triggers its BTB prefilling mechanism that requires bringing the blocks into the L1-I cache and then, feeding the pre-decoder with the arriving instructions to extract the missing branches. After resolving the BTB miss, the basic block is fed into the FTQ and in the next step, it is consumed by the fetch engine. Consequently, these blocks hit in the cache, resulting in a high miss coverage; however, the fetch engine is stalled for a considerable amount of time to resolve the BTB miss, which hurts performance. This observation also corroborates a similar study~\cite{ansaridivide}. PIF has better miss coverage as compared to MANA which is also translated to its better speedup. Moreover, it has a higher overprediction since it has a higher prefetching depth as compared to MANA.

 
\subsubsection{Instruction Prefetching Abilities}
\label{sec:instruction_prefetching_abilities}

\begin{figure}[h]
	\centering
	\includegraphics[width=1.0\textwidth]{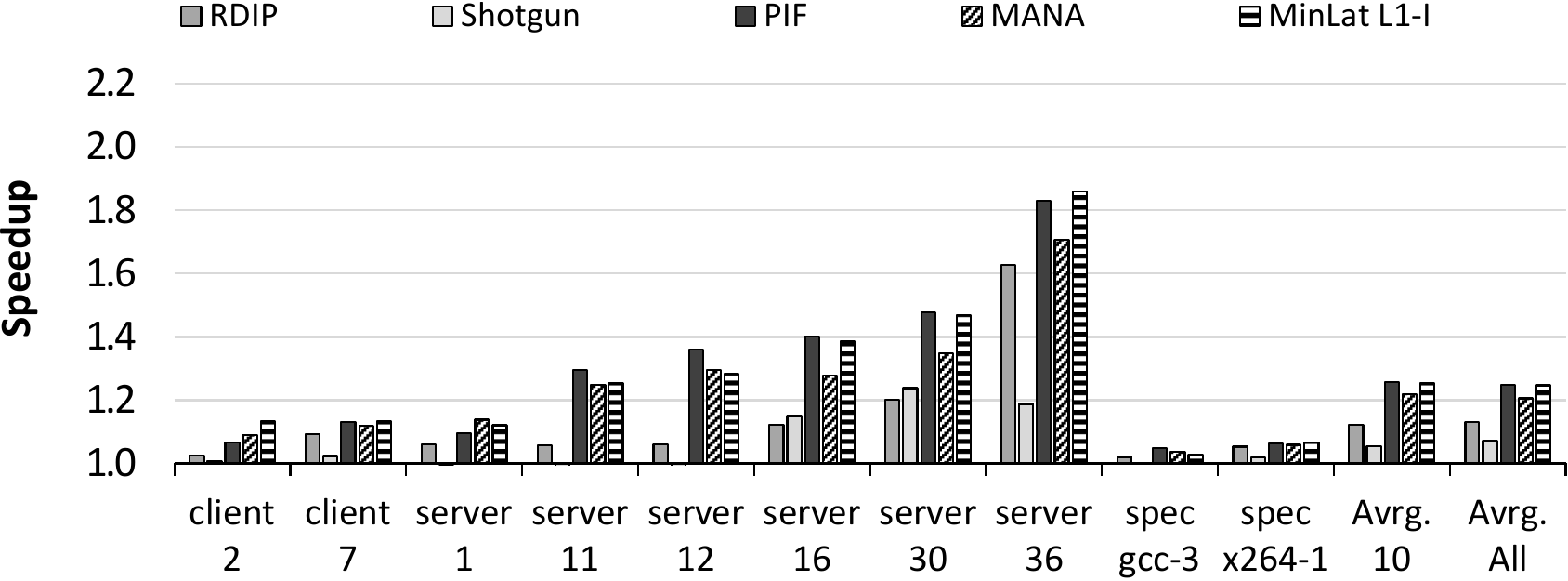}
	\caption{Speedup of competing proposals when BTB prefilling is disabled.
		\label{fig:instruction_prefetching_abilities}}
\end{figure}

In prior experiments, all competing designs could prefill the BTB to gain performance by eliminating the BTB misses. In this section, we compare them solely based on their instruction prefetching abilities. In other words, all designs have a 2 K-entry BTB that will not be prefilled. As BTB prefilling is vital for Shotgun because of its very small C-BTB, we provide two separate BTBs to evaluate Shotgun in this section: a 2 K-entry BTB is used to drive the fetch engine that will not be prefilled similar to its competitors, and Shotgun's BTBs that are only used for L1-I prefetching and benefit from BTB prefilling. Results are shown in Figure~\ref{fig:instruction_prefetching_abilities}. Comparing \textit{MinLat L1-I} in this figure and \textit{MinLat L1-I + Perfect BTB} in Figure~\ref{fig:speedup_state_of_the_arts}, we can see how much eliminating BTB misses is crucial to have a high performance instruction supply. Moreover, we see Shotgun offers near the same level of performance when we compare its results in Figures~\ref{fig:speedup_state_of_the_arts} and \ref{fig:instruction_prefetching_abilities}. It means that a conventional 2 K-entry BTB that does not benefit from BTB prefilling is as strong as Shotgun's BTBs where its small C-BTB is aggressively prefilled, in terms of driving the fetch engine. RDIP, Shotgun, PIF, and MANA offer 13, 7, 25, and 21\% speedup where MinLat L1-I offers 25\% that is very close to PIF. It can be seen that for some traces like \textit{server 12}, PIF outperforms \textit{MinLat L1-I}. It is because a powerful instruction prefetcher can completely hide the miss latency while the \textit{MinLat L1-I} faces some delays to process the fetch requests in the lower levels of the memory hierarchy.

\subsection{MANA with a Smaller Cache Size}

In this section, we use MANA to prefetch for a smaller cache-size. This study has two goals. First, by decreasing the size of the L1-I cache, the number of misses increases, and hence, puts more pressure on the prefetcher. Moreover, if the prefetcher still offers good speedup when it is used with a smaller cache-size, we can reduce the instruction cache size to provide space for MANA prefetcher. However, this design increases the traffic between the L1-I and L2 caches. Moreover, we expect the L2 external bandwidth usage does not change as almost all of the prefetch requests are served by the L2. To provide quantitative analyses, we show the speedup and the L1-I and L2 external bandwidth usage when MANA is used to prefetch for a 16~KB and an 8~KB L1-I cache. By external bandwidth usage, we mean the number of fetch and prefetch requests that are sent to the lower level of the cache hierarchy to the number of fetch requests that are sent when we have used a 32~KB cache with no prefetcher. 

\begin{figure}[h]
    \centering
    \includegraphics[width=1.0\textwidth]{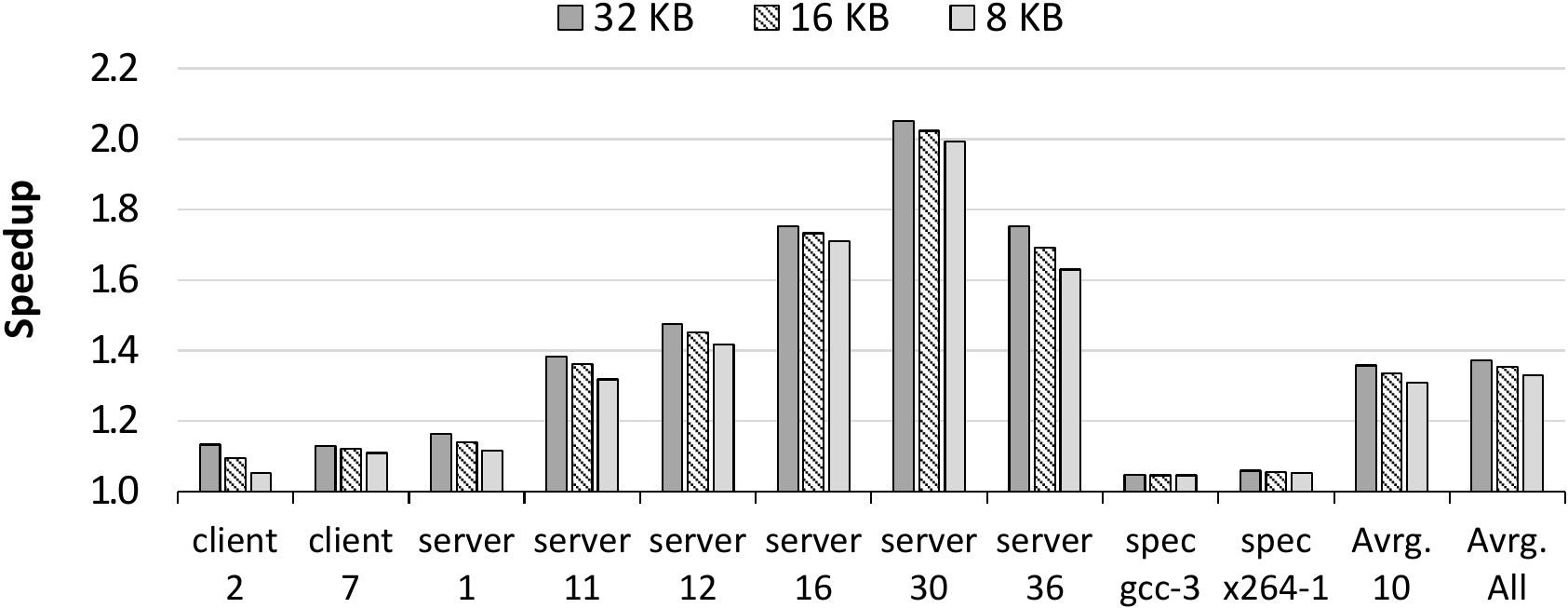}
    \caption{Speedup of MANA for various L1-I cache sizes.
        \label{fig:speedup_cachesize}}
\end{figure}

Figure \ref{fig:speedup_cachesize} shows that when we decrease the L1-I cache size, MANA still offers good speedup. The speedup is 35\% and 33\% for 16~KB and 8~KB caches, respectively. Note that MANA is designed to be independent of what is going on in L1-I caches. In other words, MANA does the same independent of the L1-I cache. The offered speedup on 16~KB cache is very close to the speedup obtained by the conventional 32 KB cache. So, we can use MANA with a 16~KB cache to avoid imposing storage overhead. This way, the design imposes no storage overhead while offers almost the same performance as MANA with a 32~KB cache.

\begin{figure}[h]
    \centering
    \includegraphics[width=1.0\textwidth]{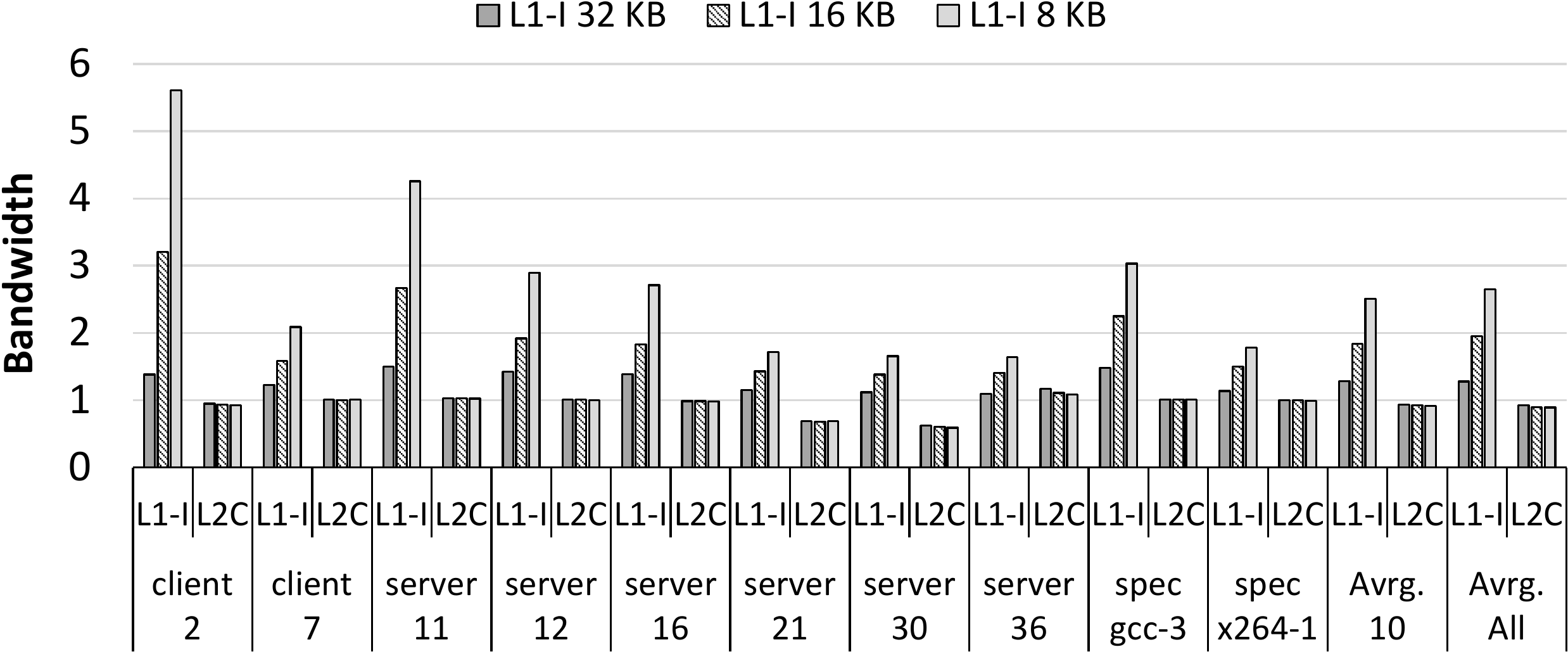}
    \caption{L1-I and L2 external bandwidth usage when MANA prefetches for various L1-I cache sizes.
        \label{fig:bandwidth_cache_size}}
\end{figure}

As expected, however, the external bandwidth usage increases by decreasing the L1-I size. Figure~\ref{fig:bandwidth_cache_size} shows that the external bandwidth usage of 16~K and 8~K L1-I caches increases by 2$\times$ and 2.65$\times$, respectively. While the L1-I external bandwidth increases, as it is the bandwidth between L1 and L2 within the chip, it is beneficial to trade it for significant performance improvement. On the other hand, the L2 external bandwidth usage does not change (slightly decreases in some cases) by decreasing the L1-I cache size. We find that almost all of the L1-I requests are served by the L2 cache (no extra traffic). Moreover, L1-I requests regularly promote the instruction cache blocks in the L2 cache to the \textit{most recently used (MRU)} position, helping them to stay a longer period in the L2 cache, resulting in lower L2 external traffic in some cases. 

        \section{Related Work}

Many pieces of prior work showed that the frontend bottleneck is a major source of performance degradation~\cite{ailamaki:dbmss, ajorpaz2018exploring, cao:detailed, hardavellas:database, keeton:performance, lo:analysis, ranganathan:performance, stets:detailed, ayers2019asmdb}. A myriad of proposals suggested prefetchers to address this problem~\cite{annavaram:call, ferdman:pif, ferdman:tifs, kallurkar:ptask, kaynak:shift, kaynak:confluence, kolli2013rdip, kumar:shotgun, kumar:boomerang, luk:cooperative, reinman:fetch, spracklen:effective, srinivasan:branch, veidenbaum:instruction, yan2008analyzing, zhang2002execution, pierce:wrong, reinman2001optimizations, chen1997instruction}. 

Some of these proposals are branch predictor-directed prefetchers and leverage the branch predictor unit to run ahead of the fetch stream to discover the missed blocks and prefetch them~\cite{chen1997instruction, reinman2001optimizations, reinman:fetch, pierce:wrong, srinivasan:branch, veidenbaum:instruction, kumar:boomerang, kumar:shotgun}. 

Discontinuity prefetcher~\cite{spracklen:effective} uses a sequential prefetcher to eliminate sequential misses and then records the remaining cache misses in a discontinuity table. Using the discontinuity table, it prefetches the discontinuity misses. Nonetheless, this prefetcher demands a large storage budget as each discontinuity entry consists of two distinct cache block addresses.

Code-layout optimization is another way to tackle the L1-I miss problem~\cite{ramirez:code, pettis:profile}. In these techniques, the program is profiled, and a control flow graph (CFG) is created. By chaining the most frequently executed control-flow changes in the CFG, the layout of the program is optimized. Code-layout optimization works well for workloads, where the control-flow changes are mostly static and can be determined at compile time.

Some pieces of prior work inserted prefetching instructions into the program code~\cite{ayers2019asmdb, kallurkar:ptask, annavaram:call, luk:cooperative, wang:guided, luk:architectural}. These proposals use an offline or online program-profiling to choose where the prefetching instructions should be added.

Confluence~\cite{kaynak:confluence} is the first proposal that offers a unified solution to address both instruction cache, and BTB misses. It used a pre-decoder to extract the branch instructions from the fetched or prefetched blocks to fill in the BTB. Confluence used SHIFT~\cite{kaynak:shift} as its instruction prefetcher; however, its idea can be applied to any other instruction prefetcher. In this work, we used Confluence's notion along with competing prefetchers to eliminate the BTB misses.

        \section{Conclusion}

Prior work used various prefetchers to eliminate instruction cache misses; however, as shown in this paper, they either do not offer the full potential or require excessive storage. We showed that prior proposals suffer from requiring a large number of prefetching records to offer their full potential and the high storage cost of these records. In this paper, we made a case that designing an effective and cost-efficient instruction prefetcher is about choosing the right metadata record and microarchitecting the prefetcher to minimize the storage requirement. Given these insights, we introduced MANA prefetcher. With only 15~KB storage, MANA offers a level of performance close to the best performing and highly storage-hungry instruction prefetcher. Moreover, MANA significantly outperforms all prior prefetchers when they have the same storage budget.

%


\bibliographystyle{ACM-Reference-Format}
\bibliography{ref}

\end{document}